\documentclass{aa}  
\usepackage{graphicx}
\usepackage{natbib}
\usepackage{amssymb}
\usepackage{amsmath}
\usepackage{amstext}
\usepackage{rotate}
\usepackage{rotating}
\usepackage{supertabular}
\usepackage{epsfig}
\usepackage{lscape}
\usepackage{enumerate}
\usepackage{float}

\bibpunct{(}{)}{;}{a}{}{,}
\usepackage{txfonts}
\newcommand{\gsim}{\hbox{\rlap{\lower.55ex\hbox{$\sim$}} \kern-.3em
\raise.4ex \hbox{$>$}}}
\newcommand{\lsim}{\hbox{\rlap{\lower.55ex\hbox{$\sim$}} \kern-.3em
\raise.4ex \hbox{$<$}}}

\newcommand{\nha}{\textsc{[N\,ii]}$\lambda$6584/H$\alpha$}
\newcommand{\sha}{\textsc{[S\,ii]}$\lambda\lambda$6717,6731/H$\alpha$}
\newcommand{\oha}{\textsc{[O\,i]}$\lambda$6300/H$\alpha$}
\newcommand{\ohb}{\textsc{[O\,iii]}$\lambda$5007/H$\beta$}

\newcommand{\nii}{\textsc{[N\,ii]}$\lambda\lambda$6584}
\newcommand{\sii}{\textsc{[S\,ii]}$\lambda\lambda$6717,6731}
\newcommand{\oi}{\textsc{[O\,i]}$\lambda$6300}
\newcommand{\ha}{H$\alpha$}

\begin{document}
   \title{VLT-VIMOS integral field
   spectroscopy of luminous  and ultraluminous
   infrared galaxies}
   \subtitle{II. Evidence for shock ionization caused by tidal forces in the
   extra-nuclear regions of interacting and merging
   LIRGs\thanks{Based 
   on observations collected at the 
   European Organisation for Astronomical Research in the Southern
   Hemisphere, Chile (ESO Programs 076.B-0479(A), 078.B-0072(A) and
   081.B-0108(A)).}}
 
  \author{Ana Monreal-Ibero\inst{1}
          \and
          Santiago Arribas\inst{2}
          \and
          Luis Colina\inst{2}
          \and
          Javier Rodr\'{\i}guez-Zaur\'{\i}n\inst{2}
          \and\\
          Almudena Alonso-Herrero\inst{2}
          \and
          Macarena Garc\'{\i}a-Mar\'{\i}n\inst{3}
          }

   \offprints{A. Monreal-Ibero}

   \institute{European Organisation for Astronomical Research in the Southern
   Hemisphere (ESO); 
              Karl-Schwarzschild-Strasse 2
              D-85748 Garching bei M\"unchen\\
              \email{amonreal@eso.org}
             \and
             Departamento de Astrof\'{\i}sica Molecular e Infrarroja
              (DAMIR), Instituto de Estructura de la Materia (IEM/CSIC);
	      c/ Serrano 121, 28996 Madrid \\ 
             \email{[arribas,colina,aalonso,jrz]@damir.csic.es}
             \and
             I. Physikalisches Institut, Universit\"at zu K\"oln
             Z\"ulpicher Strasse 77, 50937 K\"oln, Germany\\
             \email{maca@ph1.uni-koeln.de}
             }

   \date{accepted version}

 
  \abstract
   {Luminous infrared galaxies (LIRGs) are an important class
   of objects in the low$-z$ universe bridging the gap between normal
   spirals and the strongly interacting and starbursting ultraluminous infrared
   galaxies (ULIRGs). Since a large fraction of the stars in the Universe
     have been formed in these objects, LIRGs are also relevant in a
     high-$z$ context. Studies of the two-dimensional
   physical properties of LIRGs are still lacking.} 
   {We aim to understand the nature and origin of the ionization mechanisms
  operating in the extra-nuclear regions of LIRGs as a function of the
  interaction phase and infrared luminosity.} 
   {This study uses optical integral field spectroscopy (IFS) data
     obtained with VIMOS. Our analysis is based on over
     25\,300 spectra of 32 LIRGs covering all types of morphologies
     (isolated galaxies, interacting pairs, and  advanced mergers),
     and the  entire $10^{11} - 10^{12} L_{\sun}$ infrared luminosity range.
} 
 {We found strong evidence for shock ionization, with a clear
   trend  with  the dynamical status of the system.  Specifically, we
   quantified the variation with interaction phase of several
   line ratios indicative of the excitation degree. While the \nha{}
   ratio does not show any significant change, the \sha{} and
 \oha{} ratios are higher for more advanced interaction stages. 
 Velocity dispersions are higher than in normal spirals and increase with the interaction class (medians of 37, 46, and 51~km~s$^{-1}$ for class 0, 1, and 2, respectively).
 We constrained the main mechanisms causing the ionization in the
 extra-nuclear regions (typically for distances ranging from
   $\sim$0.2-2.1~kpc  to $\sim$0.9-13.2~kpc) using diagnostic diagrams. Isolated
 systems are mainly consistent with ionization caused by young
 stars. Large fractions of the extra-nuclear regions in interacting
 pairs and more advanced mergers are consistent with
   ionization caused by shocks of $v_s \lsim
   200$~km~s$^{-1}$.   
This is supported by the relation between the excitation
degree and the velocity dispersion of the ionized gas, which
  we interpret as evidence for shock ionization in
interacting galaxies and advanced mergers but not in
isolated galaxies. This relation does not show any dependence with the
infrared luminosity (i.e. the level of star formation). All
this indicates that tidal forces play a key role in the origin of the
ionizing  shocks in the extra-nuclear regions.
We also showed for the first time 
what appears to be a common  $\log$(\oha)
  - $\log (\sigma)$ relation for the extranuclear ionized gas
  in interacting (U)LIRGs  (i.e. covering the entire
  $10^{11.0}-10^{12.3}$~L$_{\odot}$ luminosity range). This
  preliminary result needs to 
  be investigated further with a larger sample of ULIRGs.
} 
   {}


 \titlerunning{VLT-VIMOS integral field spectroscopy of LIRGs and ULIRGs II.}
   
   \maketitle
%

\section{Introduction}

Luminous and ultraluminous infrared galaxies (LIRGs and ULIRGs) are
defined as those objects with an infrared luminosity of $L_{IR} = L
(8-1000 \mu \mathrm{m}) =  10^{11} - 10^{12} L_{\sun}$ and $L_{IR}
\ga 10^{12} L_{\sun}$, respectively \citep[see][for a review]{san96,lon06}.
They are systems which contain large amounts of gas and dust
\citep[e.g.][]{eva02} and which are undergoing an intense star-formation episode
in their (circum)nuclear regions 
\citep[e.g.][]{sco00,alo06}. This activity is the main cause of their
huge luminosity in about $\sim$80\% of these systems, although some
contribution from an AGN is present and even dominant in some cases
\citep[e.g.][]{gen98,ris06,far07,nar08}.

These systems usually present some degree of interaction whose
importance increases with luminosity. While the majority  of local
LIRGs can be classified as isolated spirals or interacting pairs
\citep[e.g.][]{arr04,alo06,san04}, most of the ULIRGs show signs of a clear
merging process \citep[e.g.][]{cle96,bor00,cui01,bus02,vei02}. 
While (U)LIRGs are an oddity in the local Universe, recent mid-infrared and
submillimeter surveys show how they present a strong evolution
with redshift, increasing their number by two orders of 
magnitude at $z \sim 0.8 - 1.2$ \citep{elb02}. Indeed they are the
dominant population of the infrared selected galaxies at high redshift, 
making a significant contribution to the star-formation rate density
at $0.5 < z < 2$ \citep{per05,lef05}.

The study of the ionization properties of the gas in these objects is
relevant for two main reasons. On the one hand, the ionization is
important to investigate the nature (i.e. starburst, AGN) of the
dominant source that causes the huge luminosity in the
infrared. In the optical, this has mainly been done via long-slit
observations of the nuclear regions of large samples of (U)LIRGs
\citep[e.g.][and references therein]{kim95,vei99}. These studies established trends with the luminosity and interaction stage, 
and found an increase in the 
frequency of AGN-dominated systems with luminosity. 
These results have been recently revisited using the new
  optical classifications provided by the use of \emph{Sloan Digital
    Sky Survey} (SDSS) data \citep{yua10}. They show that
  most of the (U)LIRGs previously classified as \emph{Low-ionization nuclear emission-line region}  (LINER), 
now are classified as starburst-AGN composite galaxies.
The presence of an obscured AGN has been also revealed by the
detection of ionization cones with integral
field spectroscopy (IFS) data \citep[e.g. Arp~299,][]{gar06}.

On the other hand, the ionization structure helps to understand how the
interaction/merger process as well as the release of energy and material
from the central source and/or starbursts are affecting the extended
structure of the 
galaxies in general and its interstellar medium in particular. In that sense,
the presence of Super Galactic Winds (SGWs) in (U)LIRGs has 
been suggested using emission \citep{hec90,leh96} and absorption
\citep{hec00,rup02,rup05b,rup05a} lines.  Tidally induced forces
associated with the interaction process itself have been also suggested as 
the cause for the ionization of the gas \citep{mcd03,col05}. Given
the complex structure of these systems, where the selection of a
preferential direction is specially difficult, these studies would
benefit from IFS data thanks to which it is possible to obtain
homogeneous two-dimensional spectral information.   

Using this technique \citet[][hereafter MAC06]{mon06}
have studied a sample of  six ULIRGs (nine galaxies),  and  found that wide
areas of the extra-nuclear extended regions presented line ratios
typical of LINERs according to the diagnostic diagrams of
\citet{vei87}. In addition, it was shown that the velocity dispersion
is positively correlated with the degree of ionization supporting the
idea that shocks are the main cause of the ionization in these areas.
However, these results were based on a relatively small sample, which
covered a restricted range in luminosity  ($\log(L_{IR}/L_\odot) =
12.03 - 12.40$) and interaction phase.  

In this paper we extend that study to a larger
sample of  32 systems, which cover the entire
$\log(L_{IR}/L_\odot) = 11.00 - 12.00 $ luminosity range (i.e. the
LIRGs range), and the different 
interaction types (i.e., isolated galaxies, interaction pairs, and
mergers remnants).  

The present study is part of a wider project
devoted to the study of the internal structure and kinematics of a
representative 
sample of low-redshift LIRGs and ULIRGs using optical and near-IR IFS
facilities \citep{arr08}. Specifically we used the INTEGRAL+WYFFOS facility
\citep{arr98,bin94} and the \emph{Potsdam Multi-Aperture Spectrograph},
PMAS \citep{rot05} in the Northern Hemisphere, and VIMOS 
\citep{lef03} and SINFONI \citep{eis03}  in the southern
one. The corresponding catalogs for the PMAS, INTEGRAL and
  VIMOS samples can be found in \citet{alo09}, \citet{gar09} and
 Rodr\'{\i}guez-Zaur\'{\i}n et al.  (in prep.), respectively.  
 
The paper is structured as follows: in Sect. \ref{secdos} we describe
the sample used in this work as well as 
the characteristics of the instrumental configuration
and technical details
regarding data reduction and analysis; Sect. \ref{sectres}
quantifies how the ionization degree varies with 
interaction stage and constrains the possible  mechanisms that cause the
ionization of the gas. Finally, a comparison with the previous results
for ULIRGs  and a discussion about  the origin of the ionization
produced by shocks in 
terms of the star formation and the interaction process are presented. 

Throughout the paper, a cosmology with  70~km~s$^{-1}$~Mpc$^{-1}$,
$\Omega_M =  0.3$ and $\Omega_\Lambda = 0.7$  is
assumed.


\section{The data\label{secdos}}

\subsection{The sample}

The present sample is drawn from the VIMOS IFS sample of (U)LIRGs presented in
\citet[][hereafter, Paper~I]{arr08}. Specifically, it includes all the
LIRGs listed in that paper, except  for
\object{IRAS~F10173+0828}, for which no emission lines were detected.
That implies 32 systems, of which 13 were isolated, 11
interacting systems (9 pairs and 2 triple), and  8
advanced mergers (i.e. classes 0, 1, and 2, respectively, according to Paper I
terminology). The methodology followed to perform the
  morphological classification is described below.

\begin{figure*}[!ht]
   \centering
\includegraphics[angle=0,width=0.49\textwidth]{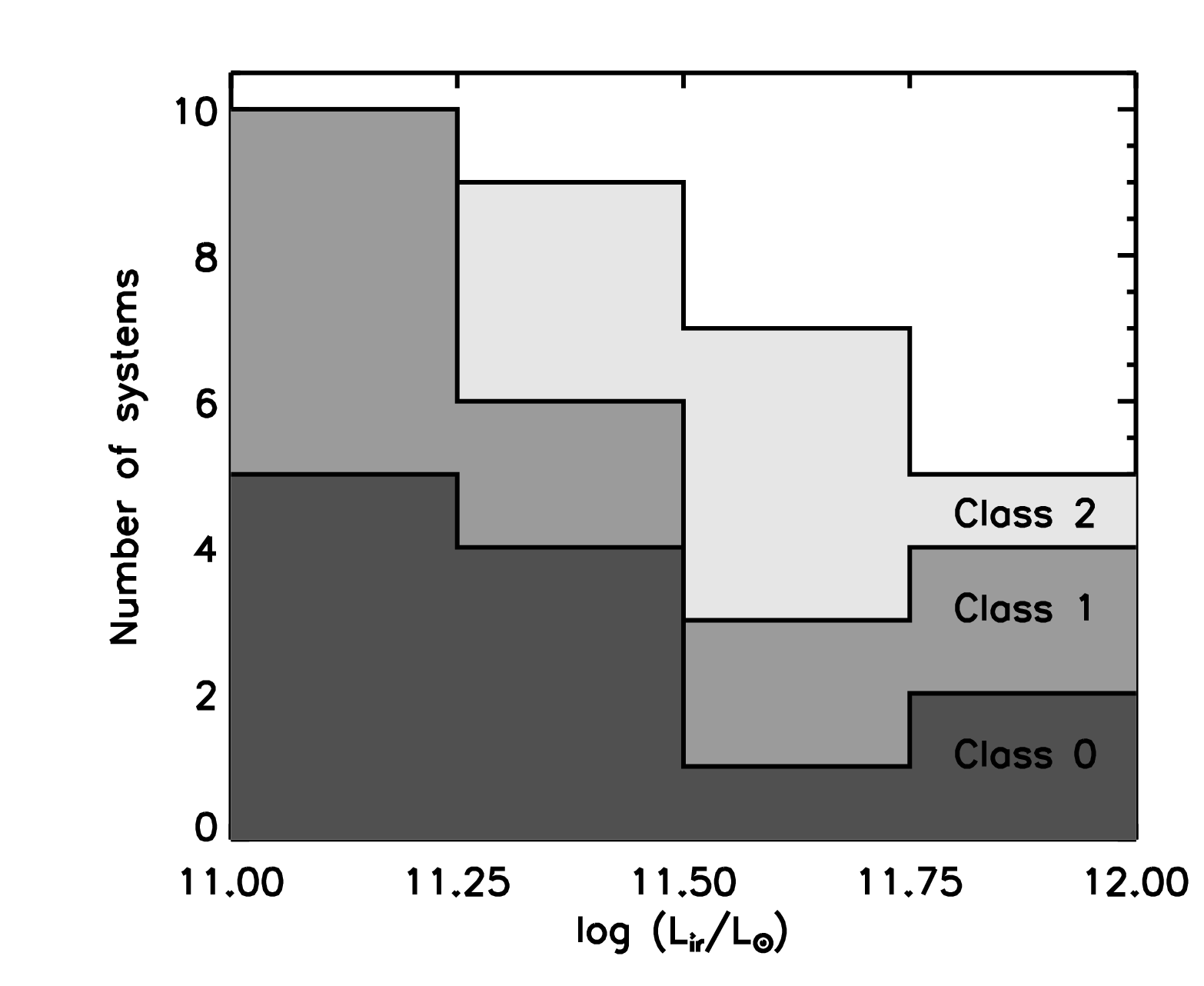}
\includegraphics[angle=0,width=0.49\textwidth]{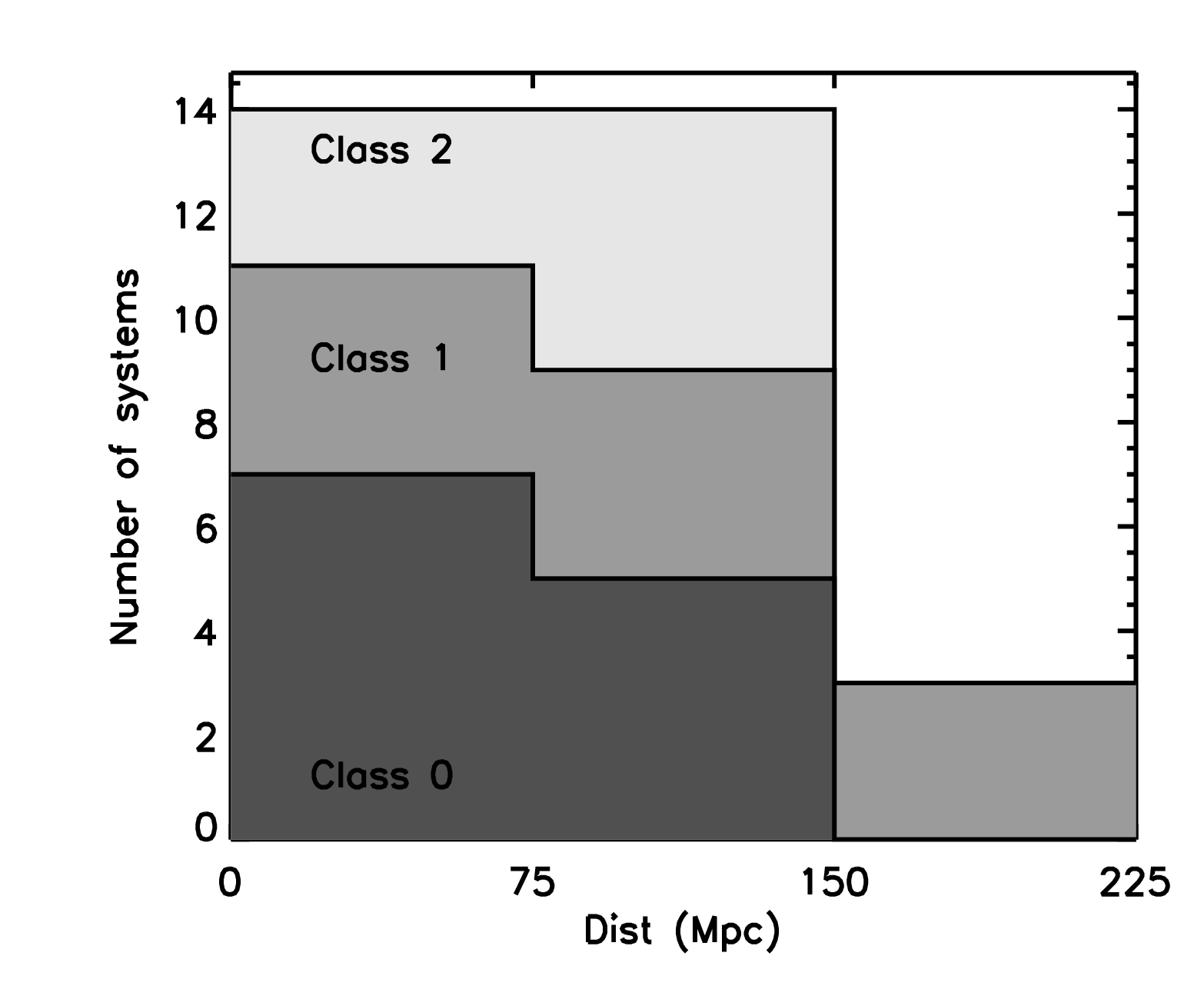}
   \caption[Histograms for the sample]
   {Histograms showing the luminosity (\emph{left}) and distance
   (\emph{right}) distributions of the systems in the sample. Light,
   intermediate and dark gray areas indicate the total number of
   isolated galaxies (class 0), interacting systems (class 1), and
   merger remnants (class 2), respectively (see text for a
     more detailed definition
   of the different morphological types).
   \label{histomuestra}}
\end{figure*}

Figure \ref{histomuestra} summarizes the distribution in luminosity,
distance, and interaction type of the systems in our final sample. 
The mean distance of the sample is 87~Mpc, which leads to a mean linear
scale of  $\sim$400~pc arcsec$^{-1}$.  Taking into account the VIMOS
spaxel size of 0\farcs67, this translates into a mean linear spatial
sampling of the source of  $\sim$270~pc spaxel$^{-1}$.   
We refer the reader to Table 1 in Paper I for the basic properties of
the individual systems.

\subsection{Morphological classification}

Our morphological/merging classification is a simplified
  version of that proposed by \citet{vei02} for ULIRGs, who divided
  their sample in  five classes (and four sub-classes). We only
  considered  three main classes (i.e. "isolated": 0, "interacting":
  1, "merger": 2) to reduce uncertainties associated with
  sorting. In particular, the different morphological classes
  considered are

\begin{itemize}

\item Class 0: Objects that appear to be single isolated
  galaxies, with a relatively symmetric morphology and without evidence
  for strong past or ongoing interaction. 

\item Class 1: Objects in a pre-coalescence phase with two
  well differentiated nuclei separated by a projected distance of D$>$1.5
  kpc. For those objects classified as 1, it is still possible to 
  identify the individual merging galaxies and their corresponding tidal
  structures due to the interaction. The limit of 1.5~kpc was
  chosen because theoretical models predict a
  fast coalescence phase after the nuclei become closer than that
  distance \citep[e.g.][]{mih96,naa06}. 

\item Class 2:  Objects with a relatively asymmetric
  morphology suggesting a post-coalescence merging phase. They may
  have two nuclei separated a projected distance of
  D$\leq$1.5~kpc. For objects classified as 2, it is not possible to
  identify the interacting galaxies individually.
\end{itemize}

For simplicity we will refer to objects of classes 0, 1, and 2
  as "isolated", "interacting", and "merger" systems. 

The classification was primarily based on the Digital Sky Survey
  (DSS) images, which are available for all the sources on the NASA
  Extragalactic database (NED). In most of the cases these images are
  sufficient to classify the objects of the present sample. However, for
  seven objects (less than 25\% of the sample), the DSS images were
  not conclusive. In these cases (\object{IRAS~08355$-$4944},
  \object{IRAS~F08520-6850}, \object{IRAS~F10038$-$3338}, \object{IRAS~F12116$-$5610},
  \object{IRAS~F13001-2339},   \object{IRAS~F17138$-$1017} and \object{IRAS~F21453$-$3511})
  high-resolution HST images were used as supplementary information,
  mainly from program ID:10582 (IP: Evans). For another six objects the
  HST images just confirm the morphological classification derived
  using the DSS images. In Figure \ref{panel} (on-line) we present the images
  used for the classification. In some cases the dynamic range is
  large and the different relevant features appear at quite different
  intensity levels, so we recommend the reader interested in a
  particular case to directly download and display the images.

In order to estimate the uncertainty associated with this 
classification process, we thrice classified the whole sample independently.
The level of agreement was
higher than 90\%. However, there were intrinsically difficult cases
for which we agreed they have an uncertain classification. These cases
are: \object{IRAS~09437+0317} (\object{IC~563}/\object{IC~564}) (1/0),
\object{IRAS~F12116-5615} (2/0), \object{IRAS~F13001-2339} (2/1),
and \object{IRAS~F17138-1017} (2/0), and they will not be
considered later when analyzing the global behavior of the different
classes. 

Further details on the morphology of these objects may be found in
Rodr\'{\i}guez-Zaur\'{\i}n et al. (in prep.).

\subsection{Observations and instrumental set-up}

A detailed description of the observations is provided in
Paper~I. Here we briefly recall the
observational set-up and  characteristics of the data.

The data were obtained in service mode with the VIMOS-IFU at VLT during
semesters 76, 78, and 81.  
We used the HR-Orange configuration, which covers the
$5250-7400$~\AA{} spectral range 
 with a resolution of 3\,400.  It provides a field-of-view of
$27^{\prime\prime} \times 27^{\prime\prime}$ with 
0\farcs67 per spatial element (\textit{spaxel})
making a total of 1\,600 spectra per pointing. Each galaxy was observed
using a 4-pointing dither pattern with a relative off-set of 2\farcs7 (i.e. four
spaxels) to minimize the effect of dead fibers and thus,
providing an effective field-of-view of about
$29\farcs7\times29\farcs7$. Details about data reduction and line fitting 
can be found in Paper I.  

\subsection{Emission line data \label{secdata}}

We used in our analysis the  [O\,\textsc{i}]$\lambda$6300, H$\alpha$,
[N\,\textsc{ii}]$\lambda\lambda$6548,6584, and
[S\,\textsc{ii}]$\lambda\lambda$6717,6730 emission lines, which were
fitted to a single Gaussian component.
For some galaxies, small areas of the H$\alpha$ emission showed
evidence for two or more kinematically distinct components. These
particular areas of double components are generally associated with
nuclear regions (i.e.  
\object{IRAS~F08520$-$6850},
\object{IRAS~F13229$-$2934}, the eastern member of
\object{IRAS~F14544$-$4255} and 
\object{IRAS~21453$-$3511})  which are not used in the present analysis (see
below). In some cases these also affect small areas in the extra-nuclear
regions (\object{IRAS~F06592$-$6313}, \object{IRAS~F07160$-$6215},
\object{IRAS~F10409$-$4556}, \object{IRAS~F13229$-$2934},
\object{IRAS~08424$-$3130}, 
\object{IC~564}, the western member of  \object{IRAS~F14544$-$4255},
the northern and central members of \object{IRAS~18093$-$5744},
\object{IRAS~F04315$-$0840}, \object{IRAS~10257$-$4338},
\object{IRAS~17138$-$1017} and \object{IRAS~21453$-$3511}).
For a given galaxy, these 
regions represent typically less than 5\% of the data. In those cases
where the two components were clearly distinguishable, the dominant
component was used in the present analysis, while in 
those of strong blending we used the results from the
one-profile fit. A more detailed analysis of the regions where 
multiple components are identified is out of the scope of the 
present paper and will be presented with the full kinematic analysis in
a future paper. 

Thus a measure of the flux in the different emission lines  as
well as one independent measurement of the central wavelength and 
FWHM (full width at half maximum) is obtained for each spectrum.  
The FWHM has been corrected from the instrumental width measured with
the strong  \textsc{[O\,i]}$\lambda$6300~\AA{} sky line, and then translated
into velocity dispersions. For the VIMOS configuration used, the
spectral resolution translates into an instrumental width
$\sigma_{ins}\sim38 \pm 8$~km~s$^{-1}$. The corresponding values for
the weaker \textsc{[O\,i]}$\lambda$6363~\AA{} sky line are   
$\sigma_{ins}\sim33 \pm 10$~km~s$^{-1}$. Therefore, the profiles of
the emission lines will be considered as resolved if their observed
profile is wider than 50~km~s$^{-1}$ ($\sigma_{obs}$). 
In general the \oi, \sii, \nii,
and H$\alpha$ emission lines define a sequence of increasing S/N and,
therefore, the \nha\ maps cover a larger area than those generated with
the other line ratios. While uncertainties in the line flux for the  
strongest emission lines (H$\alpha$) in high surface-brightness
regions are about 10\%, weaker lines like  
\sii\ and \oi\ can have large uncertainties due to the lower
S/N. However,  no lines with flux uncertainties larger than 30\% 
are used in the analysis presented in this paper. So, the typical
uncertainties in the line ratios varies from about 15\% for high S/N
lines to about 30\% for low surface brightness and when a weak line is
involved.   
 
Similarly to MAC06, we only used the extra-nuclear regions of
  our galaxies for the analysis in this paper. To do so, we excluded
  the nuclear regions, defined as those within the central
  3~spaxel$\times$3~spaxel  ($2\farcs0\times2\farcs0$) where, in
  addition to the complex line profiles mentioned above,
  contamination due to a dust enshrouded AGN could affect the results. For the
  typical distances of our sample of LIRGs, this corresponds to
  a region of about $\sim$800~pc$\times$800~pc in size. For
    the particular case of \object{IRAS~F04315-0840}, where
    the two galaxies are coalescing, the area with emission lines
    presenting double components was a bit larger than the standard
    3~spaxel$\times$3~spaxel one. Because here the one Gaussian approach did
    not properly reproduce the line profiles,  we we made an exception and also masked
    the corresponding spaxels (see Fig. \ref{mapas}).
    
The final total number of available data points, mean and standard deviation
for each individual pointing for the three line ratio used as well as the
velocity dispersion inferred from H$\alpha$ are shown in Table
\ref{galaxias}.  

   \begin{table*}[h!]
\centering
      \caption[]{VIMOS LIRG sample: For each pointing, the total
        number of data points, mean value and standard deviation for
        the different line ratios and  velocity dispersion are given. 
      \label{galaxias}} 
              \begin{tabular}{cccccccccccccccccccccccccc}
       \hline
       \hline
            \noalign{\smallskip}
Galaxy &
$\log L_{ir}^{(1)}$ &
D      &
Scale  &
\multicolumn{2}{c}{[\textsc{N\,ii}]$\lambda$6584/H$\alpha$} &
\multicolumn{2}{c}{[\textsc{S\,ii}]$\lambda\lambda$6717,6731/H$\alpha$} & 
\multicolumn{2}{c}{[\textsc{O\,i}]$\lambda$6300/H$\alpha$}  &
\multicolumn{2}{c}{$\sigma^{(2)} $ (km s$^{-1}$)}\\
 (IRAS number) &
($L_\odot$) &
(Mpc) &
(pc/$^{\prime\prime}$) &
n & $\overline x\pm\sigma (x)$ &
n & $\overline x\pm\sigma (x)$ &
n & $\overline x\pm\sigma (x)$ &
n & $\overline x\pm\sigma (x)$ \\
\hline
\multicolumn{12}{c}{Class 0}\\
\hline
F06295$-$1735       & 11.27 & 92.7  & 431 & 
  998 & 0.38$\pm$0.07 & 486 & 0.31$\pm$0.06 & 194 & 0.05$\pm$0.04 & 1118 & 33$\pm$11\\
F06592$-$6313       & 11.91 & 100.0 & 464 &
  199 & 0.82$\pm$0.32 &\ldots&\ldots&\ldots&\ldots& 266 & 54$\pm$33\\
F07027$-$6011  N    & 11.64 & 137.4 & 626 &
  342 & 0.40$\pm$0.09 &  94 & 0.30$\pm$0.06 & 92 & 0.07$\pm$0.04 & 387 & 32$\pm$11\\
F07027$-$6011  S    &  & 137.4 & 626 &
  163 & 0.61$\pm$0.17 &  74 & 0.24$\pm$0.05 & 49 & 0.03$\pm$0.02 & 177 & 72$\pm$17\\
F07160$-$6215       & 11.16 & 46.7  & 221 &
  755 & 0.92$\pm$0.61 & 189 & 0.33$\pm$0.16  & \ldots&\ldots&  888 & 69$\pm$37\\
F10015$-$0614       & 11.77 & 73.1  & 343 &
 1298 & 0.41$\pm$0.10 & 850 & 0.31$\pm$0.08  & 137 & 0.03$\pm$0.01 & 1312 & 51$\pm$13\\
F10409$-$4556       & 11.26 & 91.4  & 425 &
  928 & 0.49$\pm$0.25 &\ldots&\ldots&\ldots&\ldots& 1083 & 39$\pm$16\\
F10567$-$4310       & 11.07 & 74.6  & 350 &
 1154 & 0.44$\pm$0.15 & 270 & 0.25$\pm$0.06 & \ldots&\ldots& 1153 & 29$\pm$11\\
F11255$-$4120       & 11.04& 70.9  & 333 &
  759 & 0.58$\pm$0.28 & 173 & 0.28$\pm$0.06 & \ldots&\ldots & 782 & 33$\pm$20\\
F11506$-$3851       & 11.30& 46.6  & 221 &
  887 & 0.67$\pm$0.23 & 434 & 0.33$\pm$0.13 & 327 & 0.04$\pm$0.03 & 914 & 41$\pm$14\\
F12115$-$4656       & 11.11& 80.3  & 375 &
  724 & 0.52$\pm$0.18 & 462 & 0.31$\pm$0.08 & 274 & 0.03$\pm$0.02 & 847 & 41$\pm$12\\
F13229$-$2934       & 11.29 & 59.3  & 280 & 
  865 &  1.21$\pm$0.71 & 295 & 0.42$\pm$0.19 & 146 & 0.07$\pm$0.06 & 932 &  65$\pm$24\\
F22132$-$3705       & 11.22& 49.3  & 234 & 
 1706 & 0.41$\pm$0.08 & 1490 & 0.29$\pm$0.08 & 712 & 0.03$\pm$0.02 &1714 & 32$\pm$8\\
\hline
\multicolumn{12}{c}{Class 1}\\
\hline
F01159$-$4443       & 11.48 & 99.8  & 462 &
 558  & 0.45$\pm$0.12 & 170 & 0.29$\pm$0.06 & 97 & 0.07$\pm$0.05 & 646 & 48$\pm$20\\
ESO 297-G011       & 11.18 & 75.0  & 352 &
 1339 & 0.45$\pm$0.11 & 627 & 0.32$\pm$0.08 & 109& 0.03$\pm$0.03 &1404 & 32$\pm$10\\
ESO 297-G012       &  & 75.0  & 352 &
  223 & 0.54$\pm$0.15 &  97 & 0.38$\pm$0.10 &  80& 0.10$\pm$0.08 & 226 & 91$\pm$31\\
F06076$-$2139       & 11.67 & 165.0 & 743 &
  274 & 0.63$\pm$0.45 &\ldots&\ldots&\ldots&\ldots&313 & 40$\pm$25\\
F06259$-$4708 P1   & 11.91 & 171.1 & 769 & 
    443 & 0.48$\pm$0.24 & 221 & 0.27$\pm$0.10 & 149 & 0.04$\pm$0.04 & 550 & 81$\pm$40\\
F06259$-$4708 P2   &   & 171.1 & 769 &
    279 & 0.30$\pm$0.08 & 201 & 0.36$\pm$0.11 & 163 & 0.06$\pm$0.05 & 404 & 59$\pm$20\\
08424$-$3130        & 11.04 & 70.1  & 329 &
  241 & 0.80$\pm$0.47 & 108 & 0.38$\pm$0.15 &  56 & 0.05$\pm$0.03 & 392 & 73$\pm$32\\
F08520$-$6850       & 11.83 & 205.4 & 909 &
  196 & 0.34$\pm$0.10 & 103 & 0.39$\pm$0.10 &  39 & 0.05$\pm$0.03 & 318 & 72$\pm$31\\
IC563$^{(3,4)}$        & 11.21 & 89.0  & 415 &
  873 & 0.36$\pm$0.08 & 491 & 0.33$\pm$0.10 & 264 & 0.04$\pm$0.02 & 1007 & 38$\pm$9\\
IC564 P1$^{(3,4)}$     &  & 89.0  & 415 &
  727 & 0.42$\pm$0.12 & 163 & 0.35$\pm$0.12 &\ldots&\ldots & 843 & 33$\pm$10\\
IC564 P2 $^{(3,4)}$    &  & 89.0  & 415 &
  495 & 0.37$\pm$0.08 & 77 & 0.27$\pm$0.09  &\ldots&\ldots & 635 & 36$\pm$14\\
12042$-$3140        & 11.37 & 101.1 & 468 &
  540 & 0.58$\pm$0.23 & 246 & 0.33$\pm$0.13 & 176 & 0.08$\pm$0.06 & 611 &  85$\pm$46\\
12596$-$1529        &  11.07 & 69.0  & 324 &
  433 & 0.42$\pm$0.13 & 226 & 0.41$\pm$0.14 & 35 & 0.03$\pm$0.02 & 487 &  64$\pm$34\\
F14544$-$4255 E     & 11.11 & 68.2  & 320 &
  322 & 0.45$\pm$0.19 &  53 & 0.21$\pm$0.06 & \ldots&\ldots& 370 & 55$\pm$15\\ 
F14544$-$4255 W     &  & 68.2  & 320 & 
  356 & 0.70$\pm$0.26 & 165 & 0.68$\pm$0.27 & 87 & 0.15$\pm$0.09 & 392 & 66$\pm$24\\
18093$-$5744 N      & 11.57 & 75.3  & 353 & 
 804 & 0.41$\pm$0.12 &  528 &0.36$\pm$0.09&  105 &0.03$\pm$0.02 & 886 &  46$\pm$13\\
18093$-$5744 C      &  & 75.3  & 353 & 
 373 &0.23$\pm$0.07&  204 &0.23$\pm$0.11&  61 & 0.02$\pm$0.01 & 634 &  80$\pm$35\\
18093$-$5744 S      &  & 75.3  & 353 & 
 316 & 0.46$\pm$0.10&  185 & 0.33$\pm$0.06&   28 & 0.03$\pm$0.02& 359 &  47$\pm$16\\
\hline
\multicolumn{12}{c}{Class 2}\\
\hline
F04315$-$0840       &  11.69 & 69.1  & 325 &
  763 & 0.58$\pm$0.24 & 398 & 0.42$\pm$0.13 & 407 & 0.10$\pm$0.08 & 759 & 66$\pm$32\\
08355$-$4944        &  11.60 & 113.1 & 521 &
  260 & 0.32$\pm$0.07 & 121 & 0.25$\pm$0.06 & 108 & 0.03$\pm$0.02 & 347 & 59$\pm$20\\
F10038$-$3338       &  11.77 & 149.9 & 679 &
  172 & 0.60$\pm$0.29 & 58 & 0.44$\pm$0.18 & 42 & 0.13$\pm$0.09 & 303 & 55$\pm$28\\
10257$-$4338        &  11.69 & 40.4  & 192 &
 1698 & 0.45$\pm$0.11 &1573 & 0.39$\pm$0.13 & 1403 & 0.06$\pm$0.05 &1744 & 61$\pm$19\\
12116$-$5615$^{(3)}$ &  11.61 & 80.3  & 375 &
   90 & 0.84$\pm$0.29 & 33  & 0.34$\pm$0.13 & \ldots&\ldots & 114 & 70$\pm$15\\
13001$-$2339$^{(3)}$ & 11.48 & 94.5  & 439 &
  320 & 0.78$\pm$0.28 & 80  & 0.63$\pm$0.16 &  96 & 0.21$\pm$0.07 & 388 & 117$\pm$54\\
17138$-$1017$^{(3)}$ & 11.41 & 75.2  & 352 &
  235 & 0.53$\pm$0.15 &  87 & 0.38$\pm$0.10 & \ldots&\ldots& 257 & 63$\pm$18\\
21453$-$3511        & 11.41 & 70.0  & 329 &
 1112 & 0.61$\pm$0.41 & 476 & 0.23$\pm$0.11 &  63 & 0.04$\pm$0.05 & 1136 & 43$\pm$19\\
\hline
\end{tabular}
\begin{flushleft}
$^{(1)}$ Logarithm of the infrared luminosity, $L_{ir} =
  L(8-1000\mu$m), in units of solar bolometric luminosity, calculated
  using the fluxed in the four IRAS bands as given in \citet{san03},
  when available. Otherwise from the IRAS fluxes given in IRAS Point
  Source and Faint Source catalogs \citet{mos90}. When a system of more than
  one galaxy is observed using several VIMOS pointings, the infrared
  luminosity of the entire system is indicated in the data
  corresponding to the first galaxy of the system.\\  
$^{(2)}$ Velocity dispersion derived from the H$\alpha$ line.\\
$^{(3)}$ Classification uncertain. Not used in the analysis of systems
  grouped by interaction class.\\
$^{(4)}$ Members of \object{IRAS~09437+0317}.\\
\end{flushleft}

\end{table*}

\normalsize


\section{Results and discussion  \label{sectres}}

\subsection{General characteristics of the excitation maps \label{secmapas}}

\begin{figure*}[!ht] 
   \centering
\includegraphics[angle=0,width=.94\textwidth, clip=,bbllx=15, bblly=85,
bburx=575, bbury=810]{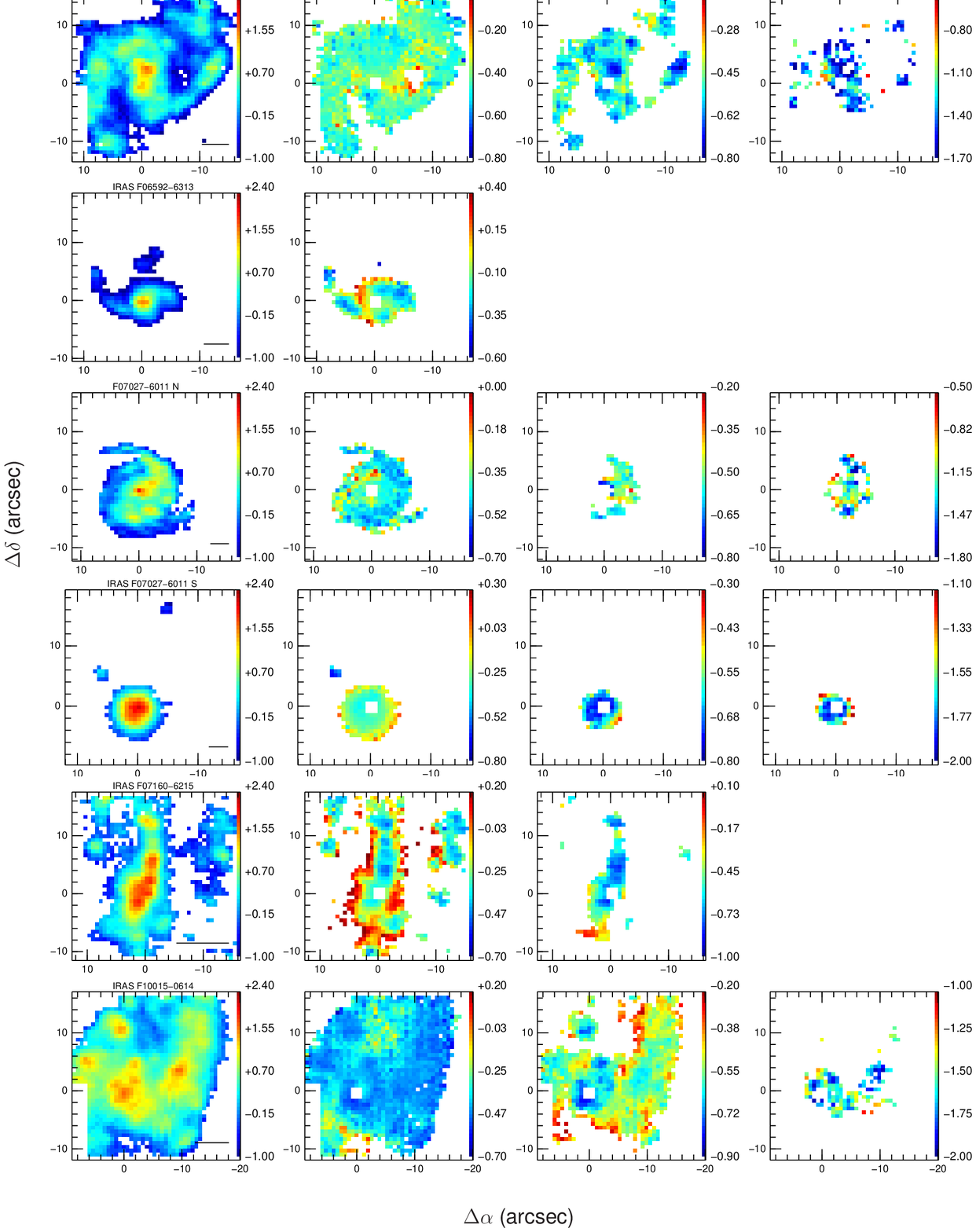}
   \caption[Maps class 0]
   {Maps for \ha\ and \nha\, \sha, and \oha\ line ratios for our LIRGs
     sample. Each row shows a VIMOS pointing. 
  The nuclear masked areas not considered in this work are
   represented in the line ratio maps as white squares. \ha\ intensity is
   displayed in arbitrary units and using a 
   logarithmic scale to better emphasize all the morphological
   features. The physical scale of 2~kpc at the distance of the galaxy
   is represented by the straight line in the bottom right corner.  Axes scales
   are in arcsec  and orientation is as usual: north up, east to the left.
   \label{mapas}}
\end{figure*}

\begin{figure*}[!ht]
   \centering
\includegraphics[angle=0,width=.94\textwidth, clip=,bbllx=15, bblly=85,
bburx=575, bbury=810]{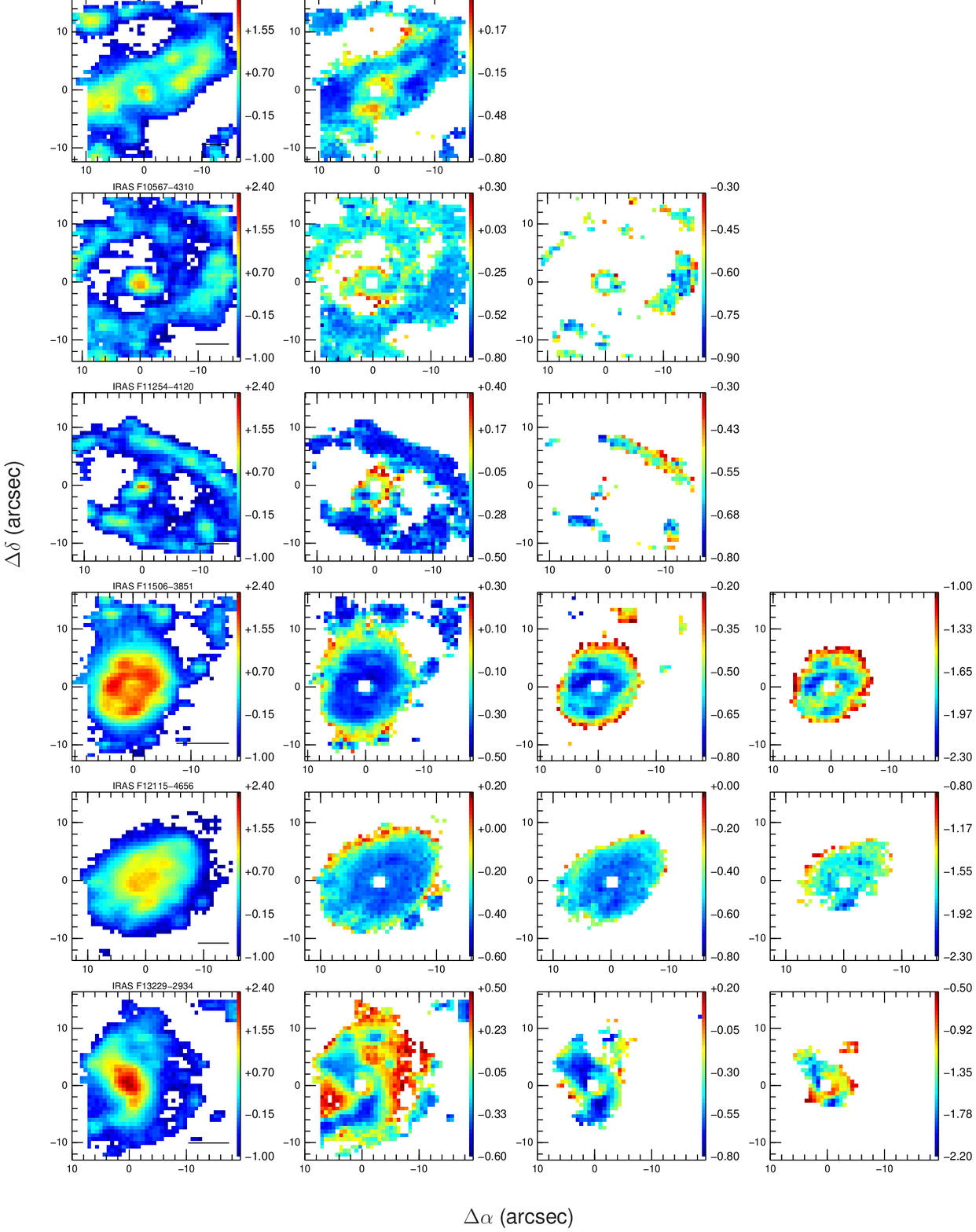}\\
\textbf{Fig. \ref{mapas}.} continued.
\end{figure*}

\begin{figure*}[!ht]
   \centering
\includegraphics[angle=0,width=.94\textwidth, clip=,bbllx=15, bblly=85,
bburx=575, bbury=810]{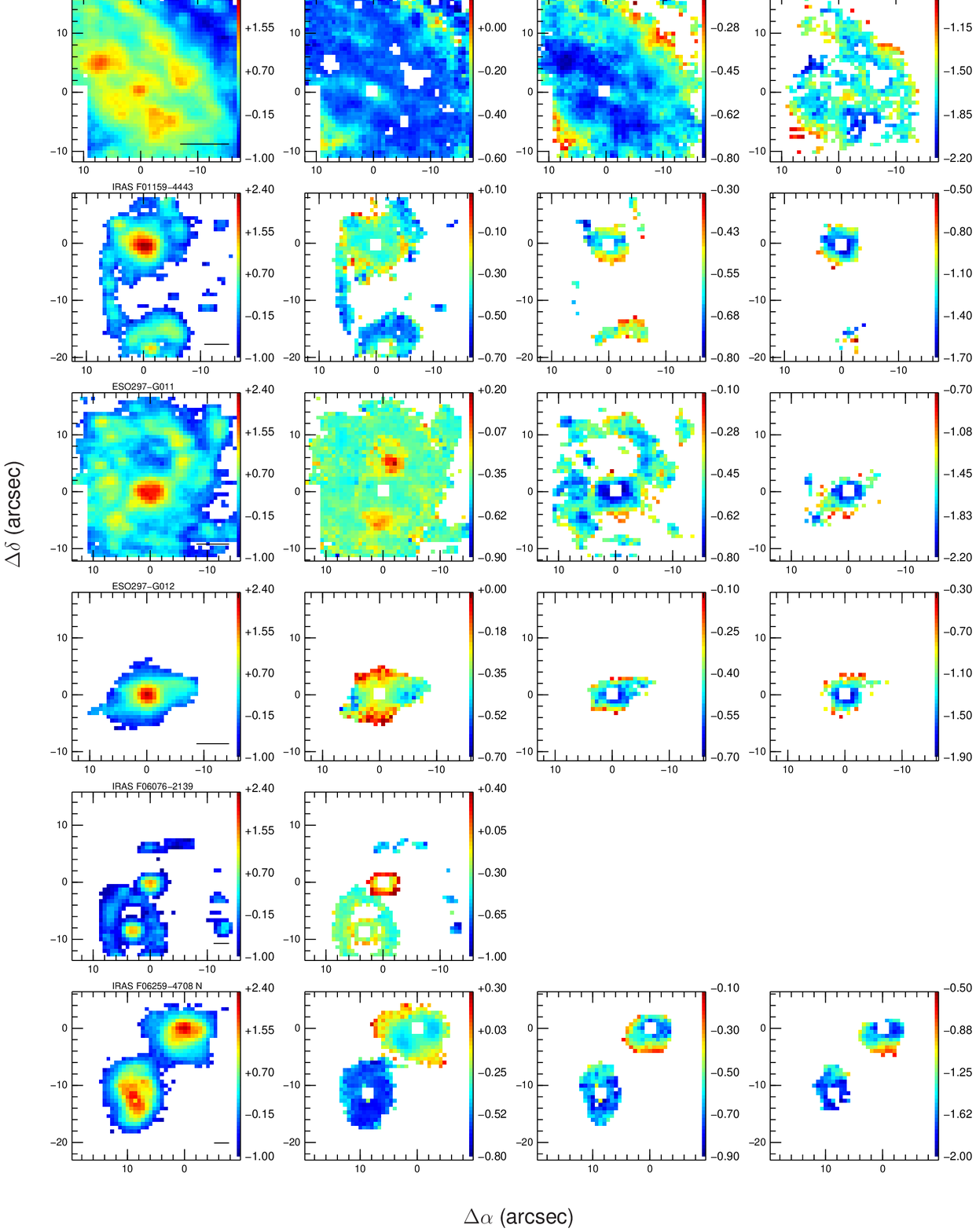}\\
\textbf{Fig. \ref{mapas}.} continued.
\end{figure*}

\begin{figure*}[!ht]
   \centering
\includegraphics[angle=0,width=.94\textwidth, clip=,bbllx=15, bblly=85,
bburx=575, bbury=810]{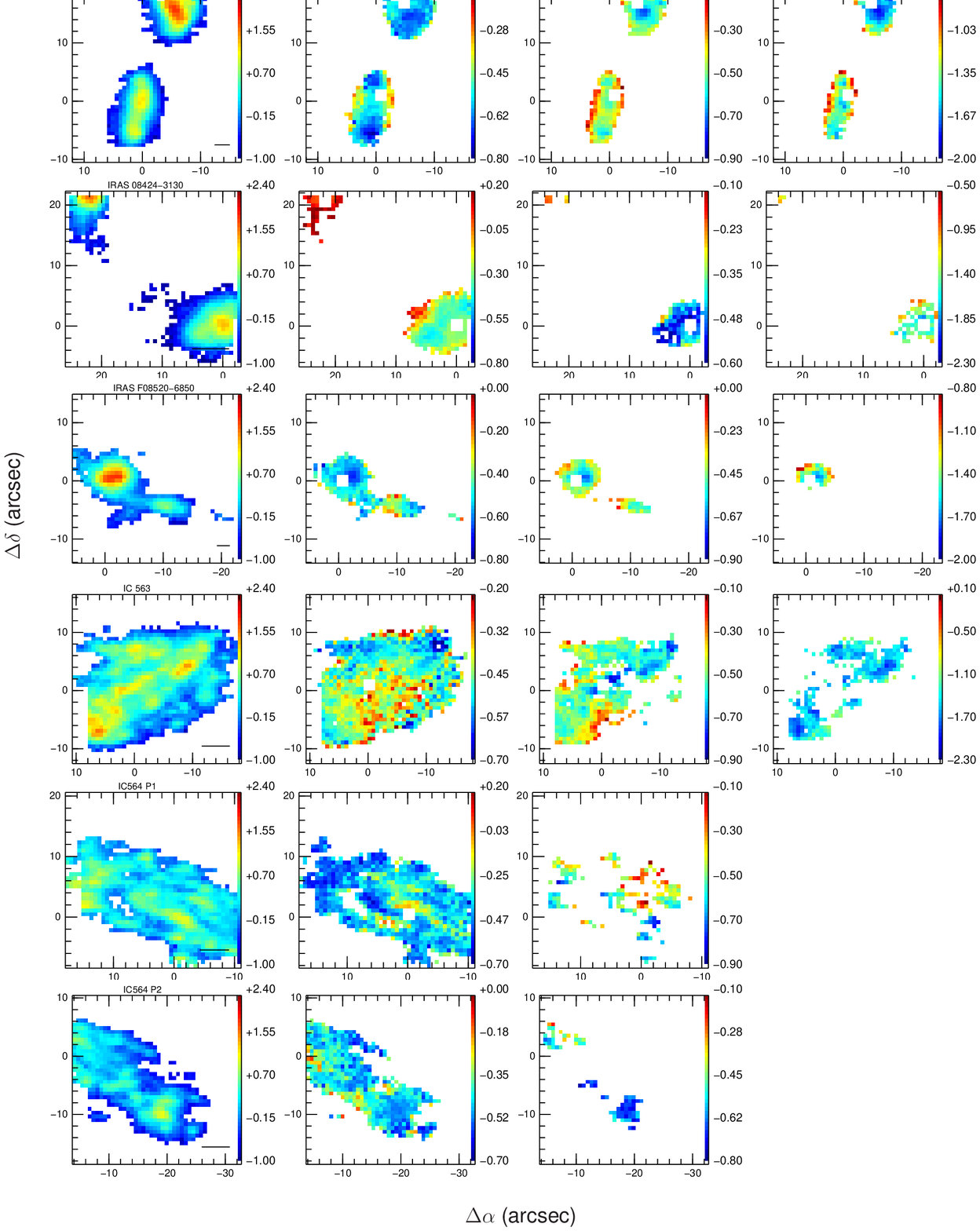}\\
\textbf{Fig. \ref{mapas}.} continued.
\end{figure*}

\begin{figure*}[!ht]
   \centering
\includegraphics[angle=0,width=.94\textwidth, clip=,bbllx=15, bblly=85,
bburx=575, bbury=810]{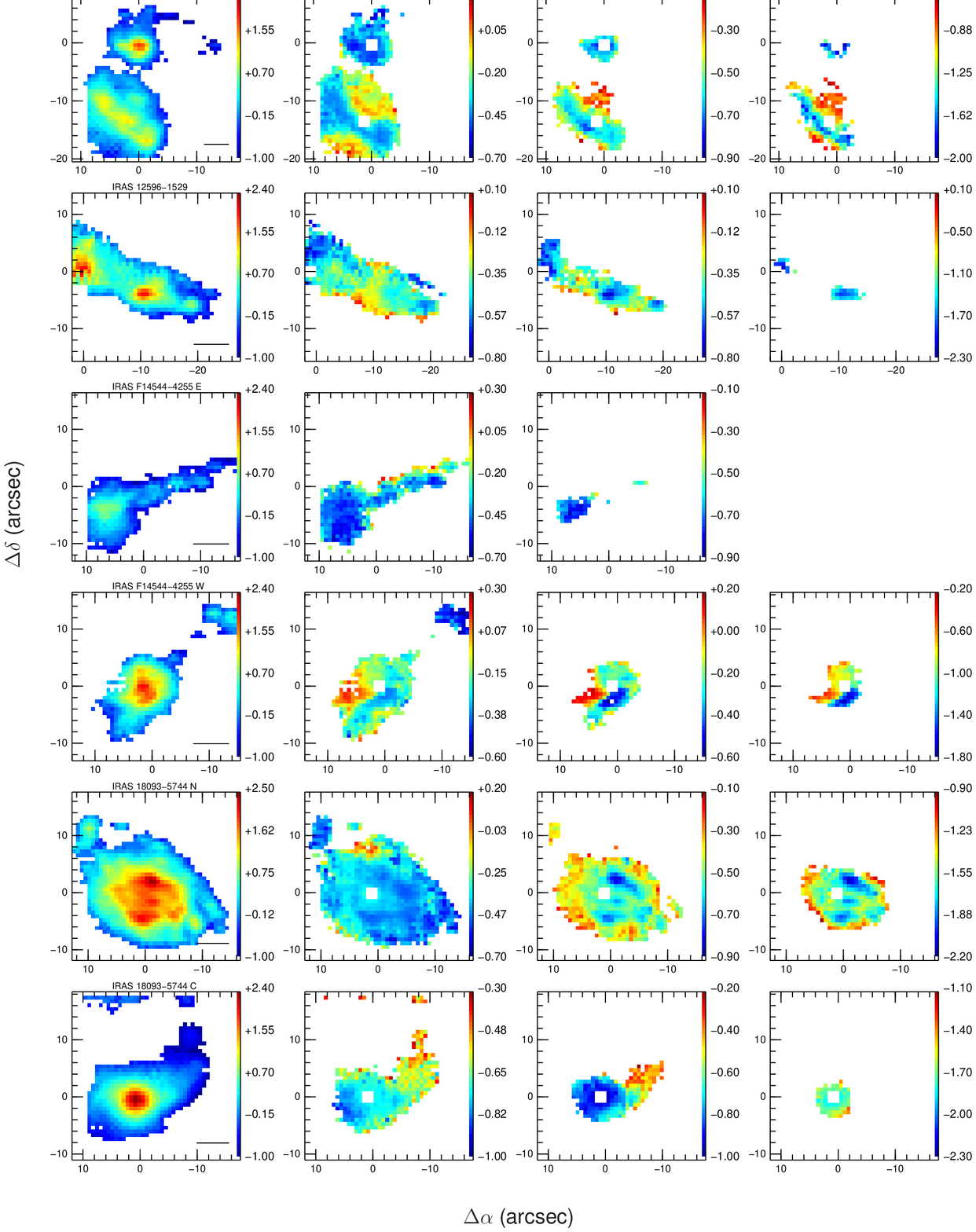}\\
\textbf{Fig. \ref{mapas}.} continued.
\end{figure*}

\begin{figure*}[!ht]
   \centering
\includegraphics[angle=0,width=.94\textwidth, clip=,bbllx=15, bblly=85,
bburx=575, bbury=810]{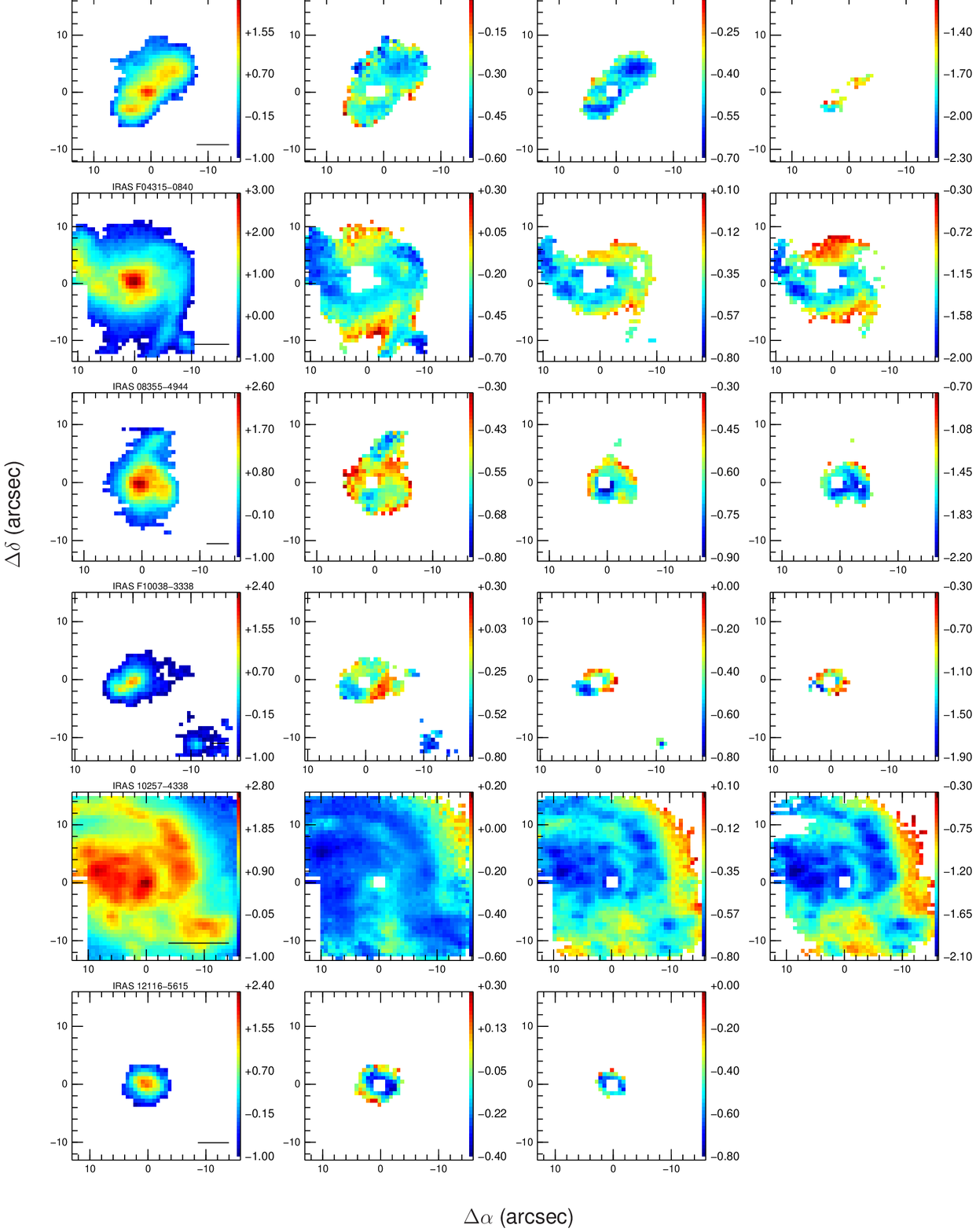}\\
\textbf{Fig. \ref{mapas}.} continued.
\end{figure*}

\begin{figure*}[!ht]
   \centering
\includegraphics[angle=0,width=.94\textwidth, clip=,bbllx=15, bblly=300,
bburx=575, bbury=690]{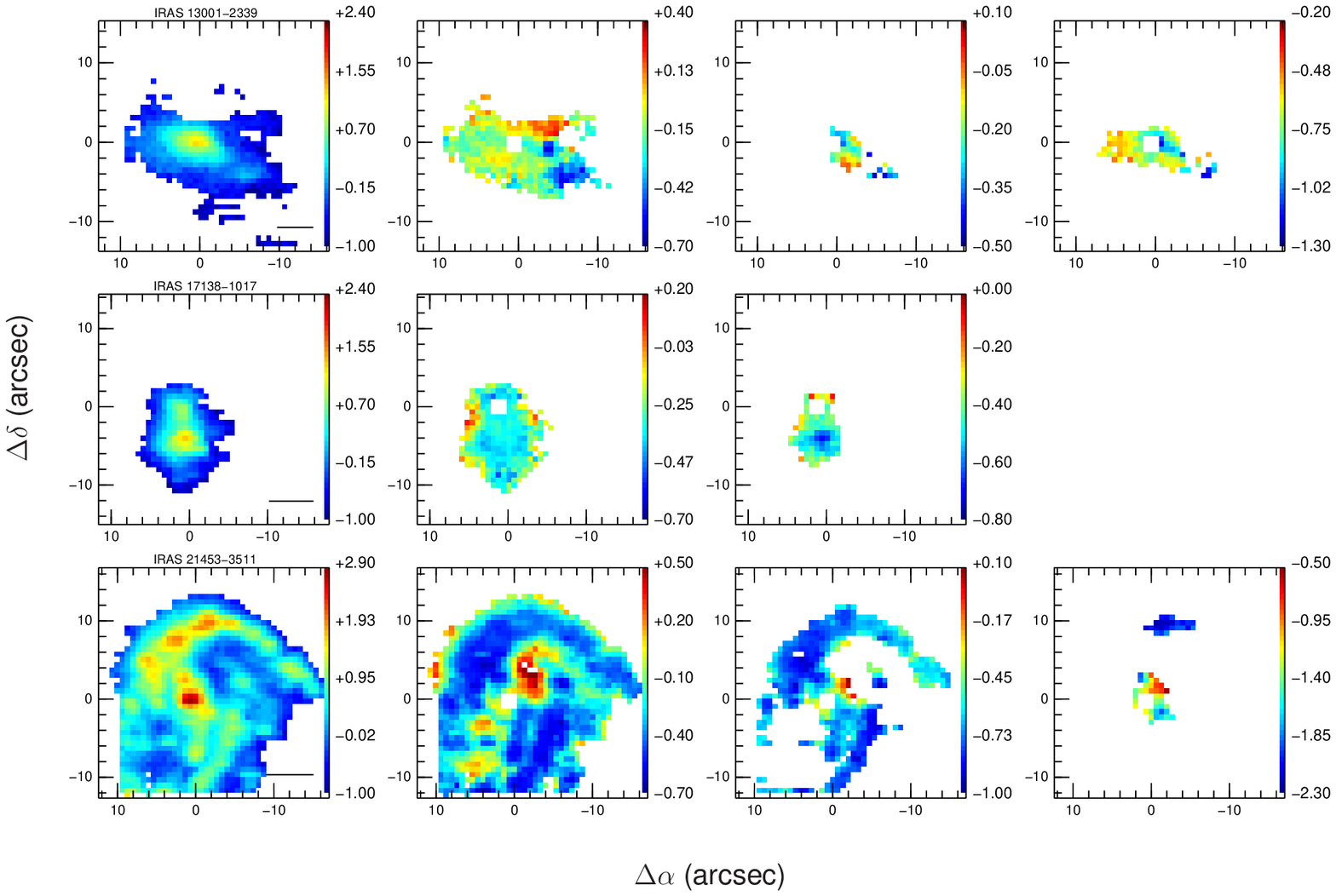}\\
\textbf{Fig. \ref{mapas}.} continued.
\end{figure*}

Figure \ref{mapas} presents the \nha, \sha, and \oha\ line
ratio maps for each individual VIMOS pointing. Although these maps
only represent one axis of the 
classical BPT diagrams \citep{bal81,vei87}, they are good
tracers of the excitation when studying the general properties 
of the extra-nuclear ionized regions of LIRGs. In fact, in the X-\ohb\
diagnostic diagrams (where X is any of the above line 
ratios),  the boundaries distinguishing  regions
mainly photo-ionized  by stars (i.e. \ion{H}{ii}-like) and those of LINER type
are nearly vertical for the expected values of $\log$(\ohb) in
these extra-nuclear regions ($\sim-0.2$, see below).  Figure 2 also shows 
the H$\alpha$ surface brightness 
maps. These maps are presented here mainly as a reference for the
excitation maps, and they will be discussed in detail  elsewhere
(Rodr\'{\i}guez-Zaur\'{\i}n et al. in prep.).  The different area covered by the
excitation maps is a direct consequence of the different
signal-to-noise (S/N) of the 
involved lines. The \nha{} map encompasses the largest  area in all the
cases, reaching areas of low H$\alpha$ surface brightness where no good S/N
data are available for the \oi\ and \sii\ lines.

The star-forming regions of relatively low excitation are well
  identified in the maps. These are generally high H$\alpha$ surface
  brightness regions  associated with large scale structures like spiral
  arms (e.g. \object{IRAS~F13229$-$2934}), rings
  (e.g. \object{IRAS~F11255$-$4120}),  or tidal tails 
  (e.g. \object{IRAS~06076$-$2139}). However, knots of star
  formation are also found in isolated external regions
  (e.g. \object{IRAS~F12115$-$4656}, \object{IRAS~F07027$-$6011}~S).
  In some cases these external regions define a chain suggestive of
  star formation along
  tidal tails  (e.g. \object{IRAS~F11506$-$3851},
  \object{IRAS~06076$-$2139}, Paper I).  

Galaxies in double or triple systems may have quite different excitation
  conditions, indicating that the interstellar medium of the individual
  objects have different properties.  For instance, while the external
  regions of the central and southern component of
  \object{IRAS~F06259$-$4708} are dominated by star formation, for the
  northern galaxy these regions show higher excitation (similarly for
  \object{IRAS~F06076$-$2139}).   

Interestingly the external regions associated to diffuse 
  low surface brightness H$\alpha$ emission have relatively high
  excitation.
In these regions the line ratios could, in principle, be affected by an
  underlying absorption spectrum.  Indeed, preliminary results for our PMAS
  sample of LIRGs \citep{alo09} show that for spectra with
  $EW($H$\alpha)\lsim10$~\AA, the line ratio can decrease $\sim$0.2~dex
  once the contribution of an old stellar population has been taken
  into account \citep{alo10}. In order to quantify the effect of a possible
  component in absorption, we created maps 
  including a correction of $EW_{abs}=2$~\AA. \sha\ and
  \oha\ line ratios are affected by about 0.2~dex for only a handful
  ($\lsim2$\%) of spaxels, and by less than  0.1~dex, for  most of the data.
  For the \nha\ line ratio, this effect could reach $\sim0.3$~dex in
  the very low surface brightness areas, but does not change in any
  case the observed ionization structure that we describe here.
These regions of low H$\alpha$ surface brightness and
  relatively high excitation are found in all kind of objects:
  structured systems with 
  rings (\object{IRAS~10567$-$4310}) and  spiral
  (\object{IRAS~F04315$-$0840}) morphologies,  highly disturbed 
  systems (the central member of \object{IRAS~F07160$-$5744}), and
  galaxies  with 
  relatively round/ elliptical  shape (\object{IRAS~F11506$-$3851}). In
  objects with round morphologies this transition 
  from low to high excitation translates into the appearance of
  external rings of high excitation (\object{IRAS~F11506$-$3851},
  \object{IRAS~F12115$-$4656}). The transition 
  from low to high excitation is also observed in the triple system
  \object{IRAS~18093$-$5744} where an extension (tidal tail?) from the
  main body of the central galaxy towards the northern galaxy is identified. 
In some cases the high-excitation regions are found close to the
nucleus and/or along preferential directions, and defining structures
suggestive of cones. Some examples are \object{IRAS~F10409$-$4556},
\object{ESO~297$-$G012} or the southern member of \object{IRAS~12042$-$3140}.

For the rest of the paper we will focus on the study of the
extra-nuclear excitation conditions using the presented line ratios as
well as the velocity dispersions.

\begin{figure*}[!ht]
   \centering
\includegraphics[angle=0,scale=.95, clip=,bbllx=45, bblly=180,
bburx=560, bbury=670]{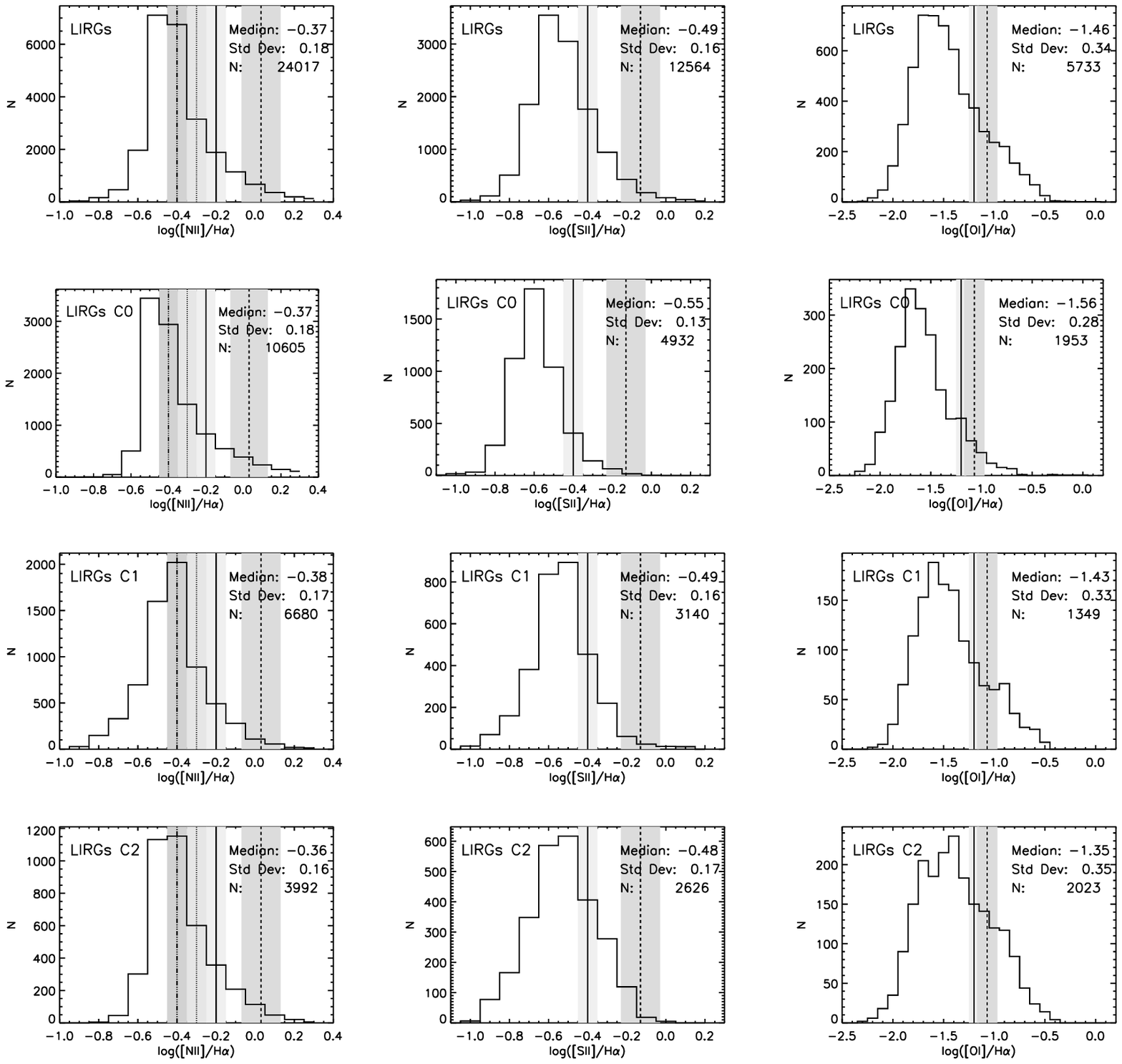}
   \caption[Emission line ratio distributions]
   {Distributions of the different LIRG subsamples according \nha,
   \sha, and \oha\ (first, second and third columns, respectively).
  Distributions are presented for the entire LIRG sample (1$^{st}$
   row), for class 0 / isolated galaxies (2$^{nd}$ row), class 1 / interacting
   (3$^{rd}$~row), and class 2 / advanced mergers (4$^{th}$
   row).
The different vertical lines indicate the boundaries between
   ionization by stars and 
   by other mechanisms proposed in different works and assuming a
   $\log$([O\,\textsc{iii}]$\lambda$5007/H$\beta$) = $-0.2$. The code used
   is: continuous line - \citet{vei87}; dotted line -
   \citet{kau03}; dashed line - \citet{kew01b}; three dots/dashed line
   - \citet{sta06}. The gray bands show the variation of a
     given boundary when the assumed  \ohb\ value changes the full
     range observed in the Alonso-Herrero et al. LIRGs sample
     (i.e. -0.6,0.1). 
   \label{histocociA}}
\end{figure*}



   \begin{table*}[h!]
\centering
\small
      \caption[]{Median, standard deviation, and number of data points
      for the distributions of the different groups and line ratios.}  
              \begin{tabular}{cccccccccccccccccccccccccc}
       \hline
       \hline
            \noalign{\smallskip}
Group &
\multicolumn{3}{c}{$\log$(\nha)} &
\multicolumn{3}{c}{$\log$(\sha)} & 
\multicolumn{3}{c}{$\log$(\oha)}  \\
 &
Median & Std Dev & N &
Median & Std Dev & N &
Median & Std Dev & N \\
\hline
All     & $-0.37$ & 0.18 & 24\,017 & $-0.49$ & 0.16 & 12\,564 & $-1.46$ & 0.34 & 5\,733\\
Class 0 & $-0.37$ & 0.18 & 10\,605 & $-0.55$ & 0.13 & 4\,932  & $-1.56$ & 0.28 & 1\,953\\
Class 1 & $-0.38$ & 0.17 & 6\,680  & $-0.49$ & 0.18 & 3\,140  & $-1.43$ & 0.33 & 1\,349\\
Class 2 & $-0.36$ & 0.16 & 3\,992  & $-0.48$ & 0.17 & 2\,628  & $-1.35$ & 0.35 & 2\,023\\
\hline
\end{tabular}
\label{estadishistogramas}
\end{table*}

\subsection{Dependence of the excitation conditions of the
       extra-nuclear ionized gas with the interaction class}

In order to investigate possible changes in the excitation properties
of the ionized gas as a consequence of interactions and mergers, the
distribution of the emission line ratios for the entire sample of
LIRGs was divided into three main groups according to their
morphology: isolated, interacting, and
merger remnants. The \nha, \sha, and \oha{}  emission
line ratio distributions for these classes of galaxies as well as for
the entire sample are 
shown in Fig. \ref{histocociA}. Galaxies with dubious
  classification (see Table \ref{galaxias}) were excluded from the
  histograms or the individual groups.
The number of  data points
as well as the median and standard deviation of these distributions
are indicated in Table \ref{estadishistogramas}.
We also indicate in each panel the boundaries between ionization
caused by stars and by other mechanisms in the
  \citet[][hereafter BPT]{bal81}
  diagrams according to the
  boundaries proposed by different authors.
The original boundaries  \citep{vei87} were empirically 
determined using a sample that includes extragalactic \ion{H}{ii}
regions, and nuclear (or integrated) data for starburst galaxies and
different kinds of active galaxies (i.e. Seyfert 2, LINERs,
narrow-line radio galaxies and what would be called today
``intermediate objects'').
Despite a recent new discussion about the location of these boundaries, we
considered that they provide
valuable information because i) they are the only set of
\emph{empirical} boundaries for the \emph{three} BPT
diagnostic diagrams ii) they facilitate a possible comparison of the
line ratios presented here with those in previous works. The other
complete set of three boundaries is the one proposed by
\citet{kew01b}. Using a combination of photo-ionization and stellar
populations synthesis models they determined the extreme
cases under which line ratios can be explained via photo-ionization
caused by stars. Recently, the empirical borders associated to the
\nha\ line ratio have been up-dated by using the \emph{Sloan Digital Sky
Survey} \citep[hereafter SDSS,][]{yor00} data \citep{kau03,sta06}. There is no
similar up-date for the diagnostic diagrams involving the \sha\ and
the \oha\ line ratios. 

In order to locate the ionization type boundaries in Fig,
  \ref{histocociA} we assumed a
  $\log$([O\,\textsc{iii}]$\lambda$5007/H$\beta$)=-0.2. This is the
  median observed value  for the sample of LIRGs sample presented in
  \citet{alo09}. 
  Individual values range  between $-0.6$ and 0.1\footnote{These
    ratios have not been corrected for the underlying stellar
    absorption in H$\beta$. Such an absorption could typically decrease the
    \ohb\ line ratio by $\sim$0.0-0.2~dex \citep[see][for
      details]{alo09}.} (excluding NGC 7469, which is known to host a
  luminous Seyfert 1 nucleus that largely affects its integrated
  spectrum).  As is illustrated in Fig. \ref{histocociA} with
  gray bands, the boundaries change very little in this \ohb\ range
  (typically $\sim$0.1~dex for the empirical boundaries of
  \citet{vei87}, \citet{sta06} and \citet{kau03} and $\sim$0.2~dex for
  the theoretical boundaries of \citet{kew01b}) making this
  $\log$(\ohb)=-0.2, a reasonable assumption for establishing the 
    \emph{mean}  borders. 

The line ratio distributions for the whole sample
(Fig. \ref{histocociA}, upper row) are not symmetrical around a mean 
value, but show a significant wing towards high values. This suggests
that a fraction of regions have a 
relatively high ionization (i.e. LINER-like). 
According to the Veilleux \& Osterbrock boundaries, the percentage of data 
presenting a LINER spectrum corresponds to 
19\%, 31\%, and  35\% when using the \nha, \sha, and
\oha, respectively. 
Using the boundaries proposed by
  \citet{kew01b}, the number of spaxels that cannot be explained as
  purely ionized by stars is significantly smaller (4\%, 2\%,
    and  25\%, for the \nha, \sha, and \oha\ line ratios, respectively). 
The significant differences in the percentages when comparing
  the three line ratios could be 
due to the fact that these distributions do not come from exactly the
same regions (i.e. set of spectra). Indeed,  the \oha{} and \sha{} data
points are 
restricted to a smaller region than the \nha{} ones because the
[\ion{O}{i}] and [\ion{S}{ii}] lines have on average lower S/N than
the [\ion{N}{ii}] 
line. However, when these distributions are generated with data points
from the same regions for the three lines (not shown), we find 
similar differences between the percentages when using the
  Kewley et al. boundaries (1\%, 1\% and 13\%) and even larger ones
  when using the Veilleux \& Osterbrock boundaries (7\%, 21\% and 22\%
  for the \nha, \sha, and \oha\ line ratios, respectively).
This confirms that the  \oha{} distribution has a
higher percentage of data points with high ionization than the other
two distributions. That  the \sii\ and  \oi\ emission
is enhanced by shocks and therefore the \sha\ and, especially, \oha\
ratio are better tracers of shock-induced ionization 
 \citep[e.g.][]{dop95} suggests a significant presence of
this type of ionization in the extra-nuclear extended regions of
LIRGs. This will be explored in more detail below.    

%
\begin{figure}[!ht]
   \centering
\includegraphics[angle=0,scale=.8, clip=,bbllx=20, bblly=5,
bburx=280, bbury=205]{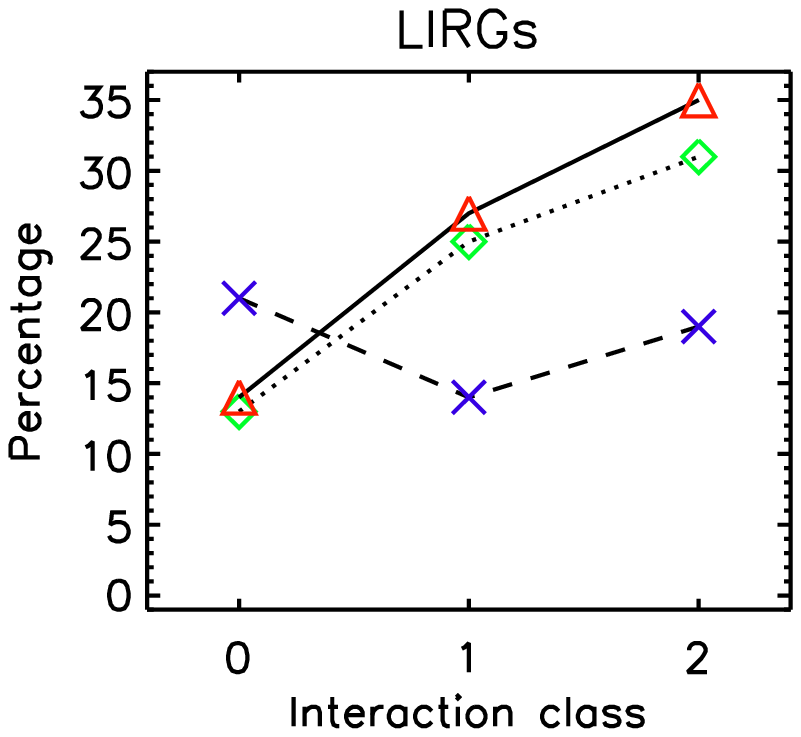}
\includegraphics[angle=0,scale=.8, clip=,bbllx=20, bblly=5,
bburx=280, bbury=205]{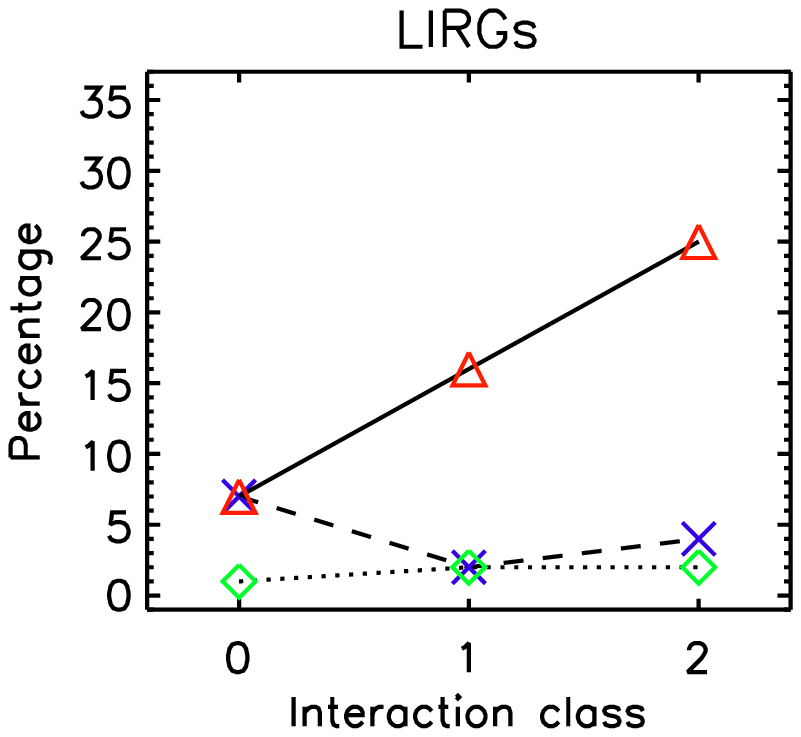}
   \caption[Median line ratio vs. Interaction class]
   {Percentage of data for the different interaction
      classes classified as LINERs using the \citet{vei87} line ratio
      boundaries (up) and \citet{kew01b} ones (down)  and assuming
      \ohb=-0.2. Measured values are indicated with
   crosses, diamonds and triangles for the \nha, \sha, and
   \oha\ line ratios respectively. Data corresponding to
   the different interaction class have been joined with dashed,
   dotted and continuous lines, respectively for an easier reading of the
   figure. \label{porcenhistogramas}}   
\end{figure}

\begin{figure}[!ht]
   \centering
\includegraphics[angle=0,scale=.8, clip=,bbllx=20, bblly=5,
bburx=280, bbury=205]{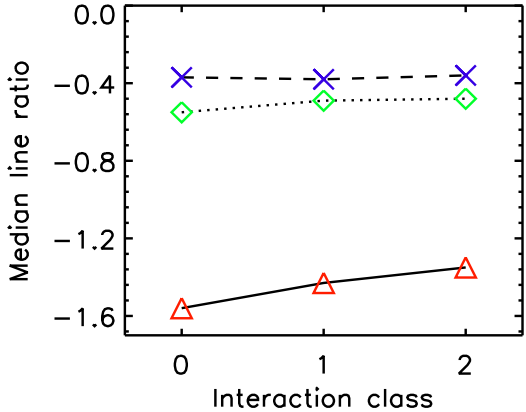}
   \caption[Median line ratio vs. Interaction class]
   {Variation of the median of the average line ratio per galaxy with
   the interaction class. We use the same line and color code as in 
   Fig.~\ref{porcenhistogramas} to distinguish among the
   emission line ratios.
\label{figresumen}}   
\end{figure}

A clear distinction emerges in the excitation properties, i.e. in the
distributions of the emission line ratios, when LIRGs are separated
according to their interaction class  (Fig. \ref{histocociA}). While
the \nha\ line ratio distributions are similar with a median ratio of $\sim
-0.37\pm0.18$~dex, typical of H\,\textsc{ii}, the
\sha\ line ratio shows on average higher values,
i.e. higher excitation, in the more dynamically perturbed systems
(class 1 and 2).  Indeed, median($\pm$standard deviation) values of
$-0.55\pm0.13$, $-0.49\pm0.16$, and $-0.48\pm0.17$ are found for
classes 0, 1, and 2, respectively. 
Similar values are found after assuming a correction for a H$\alpha$
component in absorption with an equivalent width of 2\AA. 

A more extreme change is evident in the \oha\
ratio (i.e. shock tracer) with median values of  $-1.56\pm0.28$,
$-1.43\pm0.33$, and $-1.35\pm0.35$ for classes 0, 1, and 2,
respectively. In addition, LIRGs belonging to both classes 1
  and 2 seem to have a double-peak 
distribution, with values around -1.4 (i.e. H\,\textsc{ii}-like) and
-0.9 (LINER-like). 
In all the cases, the Kolmogorov-Smirnoff test allows us to 
  reject the possibility that these distributions come from the same
  parent distribution even in the case of the \nha\ line ratio. 

Moreover, when using the Veilleux \& Osterbrock boundaries
the percentage of data points with \sha\ and \oha\ line 
ratios in the LINER range increases by factors of two and three with
the interaction class  with respect to the isolated galaxies (class
0), representing about $\sim$17\% and $\sim$33\% of the regions in
class 1 and 2 galaxies, respectively. When using the Kewley et
al. boundaries, only the percentage for the \oha\ line ratio shows a
significant increase (see Fig. \ref{porcenhistogramas}).  

As shown in Table \ref{galaxias}, the number of data
  points associated with a galaxy can range from a few tens to a few thousands,
  depending on the line ratio. Given this range of two orders of
  magnitude, one might wonder if the distributions presented in Fig.
  \ref{histocociA} are biased due to the contribution of a 
  few galaxies. To check this possibility, we generated the
  distributions for the three line ratios and the different
  morphological groups, but without
  the pointing with the largest number of spaxels. These were
  \object{IRAS~F22132$-$3705} for class 
  0, \object{ESO~297-G011} for class 1, and \object{IRAS~10257-4338}
  for class 2. In this way, the remaining pointings 
  contribute in a similar manner to the distribution (i.e. with
  several hundreds of data points for the \nha\ line ratio in most of
  the cases and with many tens - a few hundreds for the \sha\ and \oha\
  line ratios). The distributions (not shown here) were similar to those for the
  whole sample, with differences between
  the medians $\sim$0.05~dex.
Moreover, we checked how the median of the
  average line ratios per galaxy changes with the interaction
  class. This is shown in Fig. \ref{figresumen}. In this way, every pointing
  contributes with one data point per line ratio. As was found for
  the data distributions, this 
  figure shows how, while the 
  \nha\ line ratio remains constant with the interaction class, the
  \sha\ and, especially, the \oha\ line ratio increases with the
  degree of interaction.

\begin{figure*}[!ht]
   \centering
\includegraphics[angle=0,scale=.95, clip=,bbllx=45, bblly=350,
bburx=560, bbury=500]{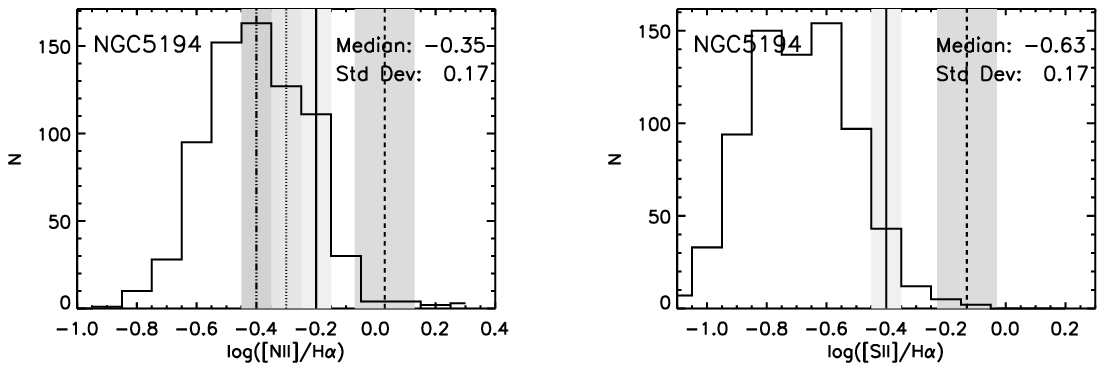}
   \caption[Emission line ratio distributions]
   {Same as Fig. \ref{histocociA} for NGC~5194 (only \nha\ and \sha\
     line ratios). \label{histoVIRUS}}
\end{figure*}

Although it would have been interesting to compare  the
  present line ratio distributions with those obtained for other kinds
  of galaxies (e.g. normal spirals, less powerful starbursts), there
  is a clear lack of data comparable to those presented here. In
  particular, a meaningful comparison requires galaxies with line
  ratios measured at similar spatial scales as those sampled here.   
  For example, line ratios derived for the SDSS\footnote{See
    \texttt{http://www.mpa-garching.mpg.de/SDSS/}.}, the largest
    extragalactic survey up to date, were derived from spectra obtained
    through 3-arcsec diameter fibers.
For galaxies at similar redshift like the present sample, this would be
  representative of the ionization in the \emph{nuclear} regions, and
  thus, not directly comparable with our analysis on the external
  areas.
For galaxies at larger distances - say, $0.04<z<0.10$ - the spatial
  sampling in the SLOAN galaxies would be much larger
  ($\diameter\sim2.4- 5.4$~kpc) and thus again not directly
  comparable.
The only galaxy with published data comparable to those presented here
is the spiral NGC~5194 \citep[also known as \emph{Whirlpool
      Galaxy,}][]{bla09}, a spiral galaxy interacting with a dwarf
  galaxy. At its distance ($\sim$8~Mpc)  
  and with the size of the VIRUS-P fibers ($\diameter\sim4\farcs3$)
  the  spatial sampling is comparable with that for our sample
  within a factor of $\sim$3. Figure~\ref{histoVIRUS} shows the
  distribution for the two available line ratios for this
  galaxy. While the \nha\ distributions for the LIRGs and NGC~5194 are
  similar, LIRGs have larger \sha\ ratios. Because this
  comparison was done only with one galaxy, this result
  needs to be revised when line ratios in the external
  areas of larger samples of spiral galaxies at similar linear
  spatial resolution become available.

\begin{figure*}[!ht]
   \centering
\includegraphics[angle=0,width=0.99\textwidth, clip=,bbllx=60, bblly=305,
bburx=508, bbury=610]{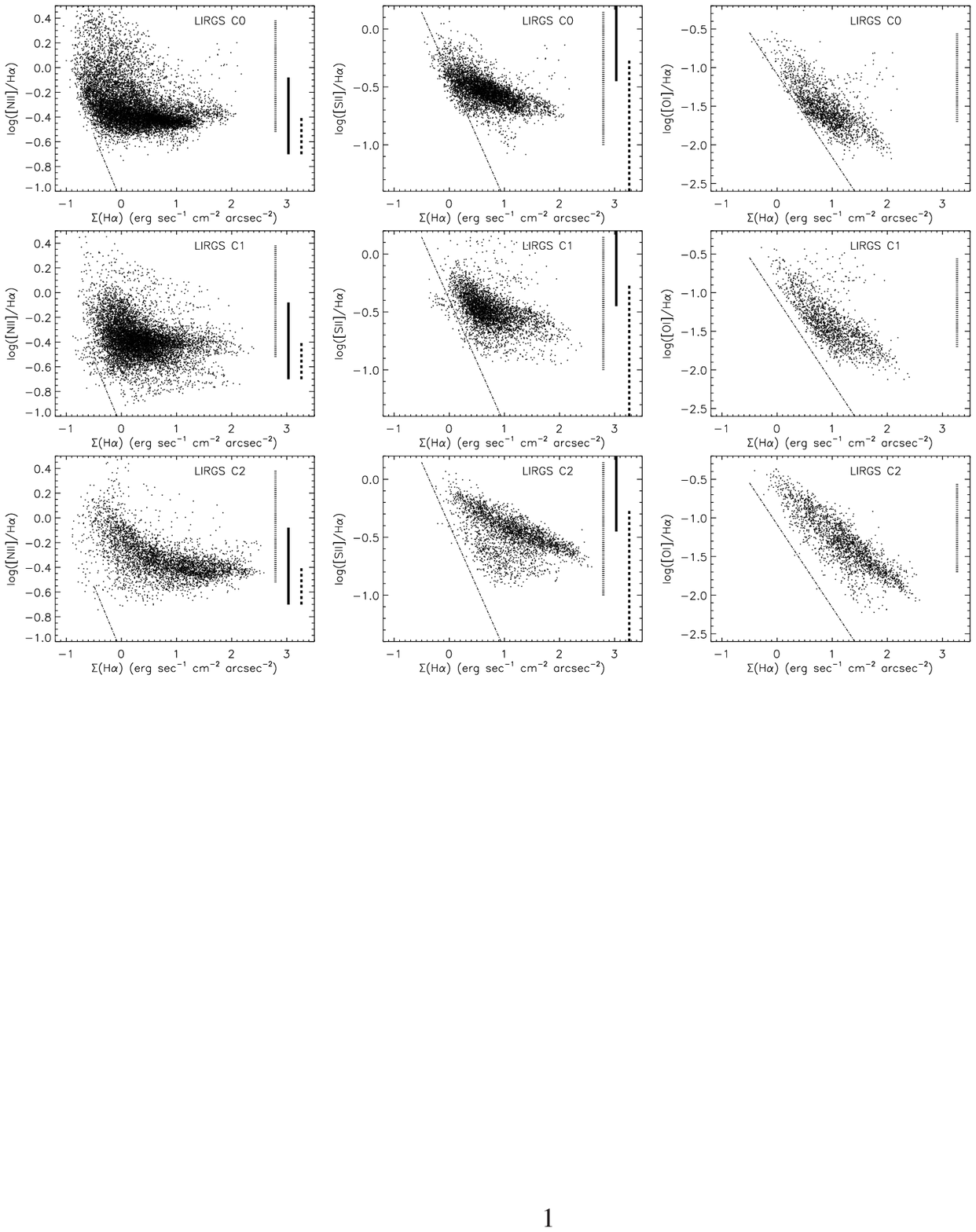}
   \caption[Luminosity vs. line ratios.]
   {Relation between the \ha\ surface brightness  and the
   [N\,\textsc{ii}]$\lambda$6584/H$\alpha$ (left),
   [S\,\textsc{ii}]$\lambda\lambda$6717,6731/H$\alpha$ (middle), and
   [O\,\textsc{i}]$\lambda$6300/H$\alpha$ (right). Units are in an
   arbitrary scaling.
   The dashed (continuous) vertical bars represent the covered line ratio ranges
   for the sample of galactic \ion{H}{ii} regions (DIG areas)
   studied  by \citet{mad06}. Note that because its
   \textsc{[S\,ii]}/H$\alpha$ line ratio only includes the
   $\lambda$6717 sulfur line, we added an offset of 0.25~dex to
   their values to take into account the contribution from the
   $\lambda$6731 line. Also the line ratio ranges for the eDIG
   areas in a sample of nine edge-on spiral galaxies is shown with dotted
   line for comparison \citep{mil03}. Dotted-dashed lines represent our
   estimated observational limits.
    \label{ionivslumi}}
\end{figure*}

\subsection{Anti-correlation between excitation and \ha\ luminosity}

Maps presented in Sect. \ref{secmapas} (Fig. \ref{mapas})
  display an increase on the 
line ratio in those areas with low \ha\ surface brightness. Similar
results have been observed in other environments, like our Galaxy
\citep{mad06,rey99}, or other spiral galaxies \citep{col01d,mil03,bla09} in
the so-called diffuse ionized gas (DIG) or warm interstellar medium (WIM).
In this section we further explore this result by looking at
the relation between the ionization degree and surface brightness
for the different interaction groups.
This is shown in Fig. \ref{ionivslumi}, where each data point
  represents the information from an individual spaxel.
As can be seen in the figure, while the \nha\ covers almost
  three orders of magnitude in \ha\ surface brightness, the \sha\ and the
  \oha\ line ratios are restricted to only about two.
Also, low values of \sha\ and \oha\ are only found at high \ha\
  surface brightness. 
We estimated our observational limit by looking at the typical values and
  uncertainties measured for our \nha\ line ratio. Then we allowed for a
  maximum uncertainty in the logarithm of the line ratios of 0.4
  and assumed that the S/N scales with the root square of the
  signal. The derived observational limit is shown in Fig.
  \ref{ionivslumi} which shows how these effects are caused by the
  sensitivity limit of the data, as was 
  pointed out in Sect. \ref{secmapas}. That is, the relatively low S/N of
  the \sii\ and \oi\ emission lines in the outer parts prevent us from
  measuring low
  \sha\ and \oha\ line ratios at low \ha\ surface brightness.
Figure \ref{ionivslumi} shows that independently of the line ratio, high
  values are found at low surface brightness as it happened in our
  Galaxy and other spiral galaxies. Moreover, typical distances where
  the high line ratios are found (from $\sim$400~pc up to $\sim$6~kpc)
  are comparable with the 
  distances where this anti-correlation between excitation and surface
  brightness has been found \citep[up to $\sim$8~kpc][]{col01d,mil03}.

The figure also shows the ranges of \nha\ and \sha\ line
  ratios measured in a sample of \ion{H}{ii} regions and DIGs areas in
  our Galaxy by \citet{mad06}, as well as for extragalactic DIG \citep{mil03}.
  Part of the \nha\ and \sha\ line ratios are compatible with those
  expected for \ion{H}{ii} regions. However, there is a large number
  of spaxels with line ratios similar to what it is observed in
  the DIGs areas.

\ion{H}{ii} regions are generally understood as photo-ionized
  by the young stellar populations within them. On the
  other hand, DIGs seem more difficult to explain only via
  photo-ionization \citep[e.g.][]{mil03}.  If they are interpreted only as
  photo-ionization areas, there is a need for extra heating
  \citep{rey99,mat00} and even with this extra heating, it is
  difficult to reach \nha$\gsim$1 line ratios. Another possibility would be
  that these line ratios are the composite effect of
  photo-ionization and shocks or turbulent 
  mixing layers \citep[TML,][]{col01d,mil03}.   
In the next section we will explore the role of these
  different  ionization mechanisms by comparing our measured line
  ratios with the predictions of the models.

\begin{figure*}[!ht]
   \centering
\includegraphics[angle=0,scale=.93, clip=,bbllx=45, bblly=235,
bburx=585, bbury=585]{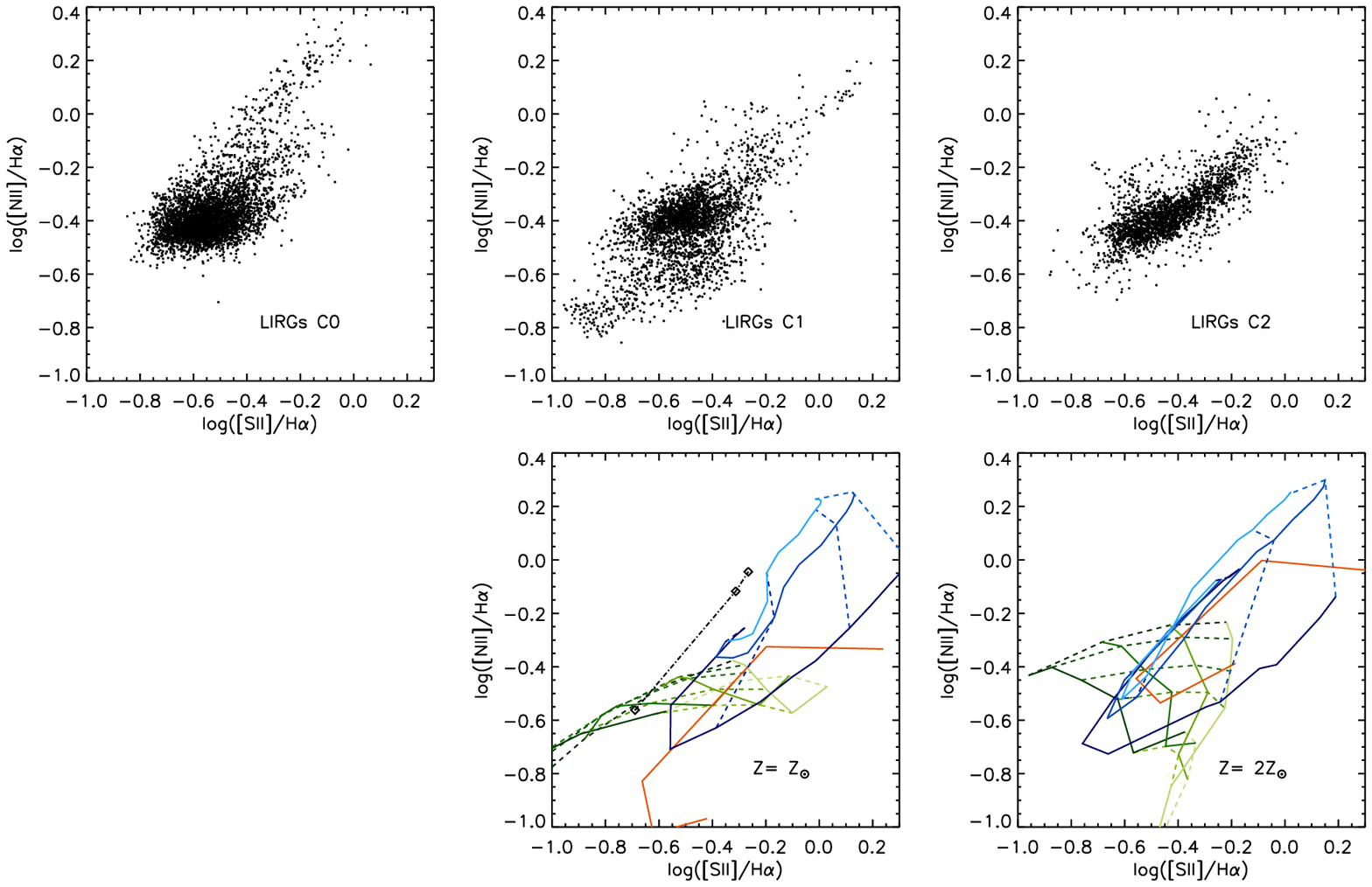}
   \caption[\sha\ vs. \nha.]
   {\sha\ vs. \nha\ diagrams. \emph{First row:} Line ratios for the
   different LIRGs groups in our sample. 
   \emph{Second row:} Theoretical models used for comparison where those in
   the middle column have solar metallicity and those in the right twice solar
   metallicity. In all panels, the empirical boundaries proposed in 
   \citet{vei87} to separate the ionization by stars or other
   mechanisms assuming a
   $\log$~([O\,\textsc{iii}]$\lambda$5007/H$\beta$) = -0.2 are shown.
   The color code for the models is:
i) \emph{Green scale:} Star-formation 
   models from \citet{dop06}. Different solid lines represent
   different $R$ parameter which vary from 0 (darkest green) to -6 (lightest
   green) while dashed lines connect points with the same age ranging
   from 0.5~Myr (darkest green) to 3.0~Myr (lightest green);
ii) \emph{Orange scale:} Dusty AGN models models from
   \citet{gro04}. Ionization parameter ranges from 0.0 (bottom) to 
   $-4.0$ (rightmost edge). Only the cases with $\alpha=-2.0$ have
   been plotted. Smaller $\alpha$ parameters typically predict higher
   \sha\ and \nha\ line ratios for a given ionization parameter.
iii) \emph{Blue scale:} Models for shocks without precursor from
   \citet{all08} for $n_e = 1 
   $~cm$^{-3}$. Solid lines indicate models for magnetic 
   parameters of $B/n^{1/2}=$ 0 (darkest blue), 2, and 4
   $\mu$G~cm$^{3/2}$ (lightest blue) while
   dashed lines joint points with velocities of 100~km~s$^{-1}$
   (darkest blue), 200~km~s$^{-1}$ (medium blue) and 300~km~s$^{-1}$
   (lightest blue). The predictions for models with shock velocities
   greater than 300~km~s$^{-1}$ 
   occupy a similar area as those for $v_s=300$~km~s$^{-1}$ in this
   diagram. Models for shocks with precursor cover a similar area in
   this diagram displaced by $\sim -0.2$~dex in both line ratios.
iv) \emph{Black dash-dotted line:} Models for turbulent mixing layers
of \citet{sla93} for $Z=Z_\odot$ and transverse velocity $v_t =
25$~km~s$^{-1}$. Average temperature for the gas increase with
   \nha. The three diamonds indicates $\log T$=5.0, 5.3 and 5.5.
    \label{figs2havsn2gha}}
\end{figure*}

\begin{figure*}[!ht]
   \centering
\includegraphics[angle=0,scale=.93, clip=,bbllx=45, bblly=235,
bburx=585, bbury=585]{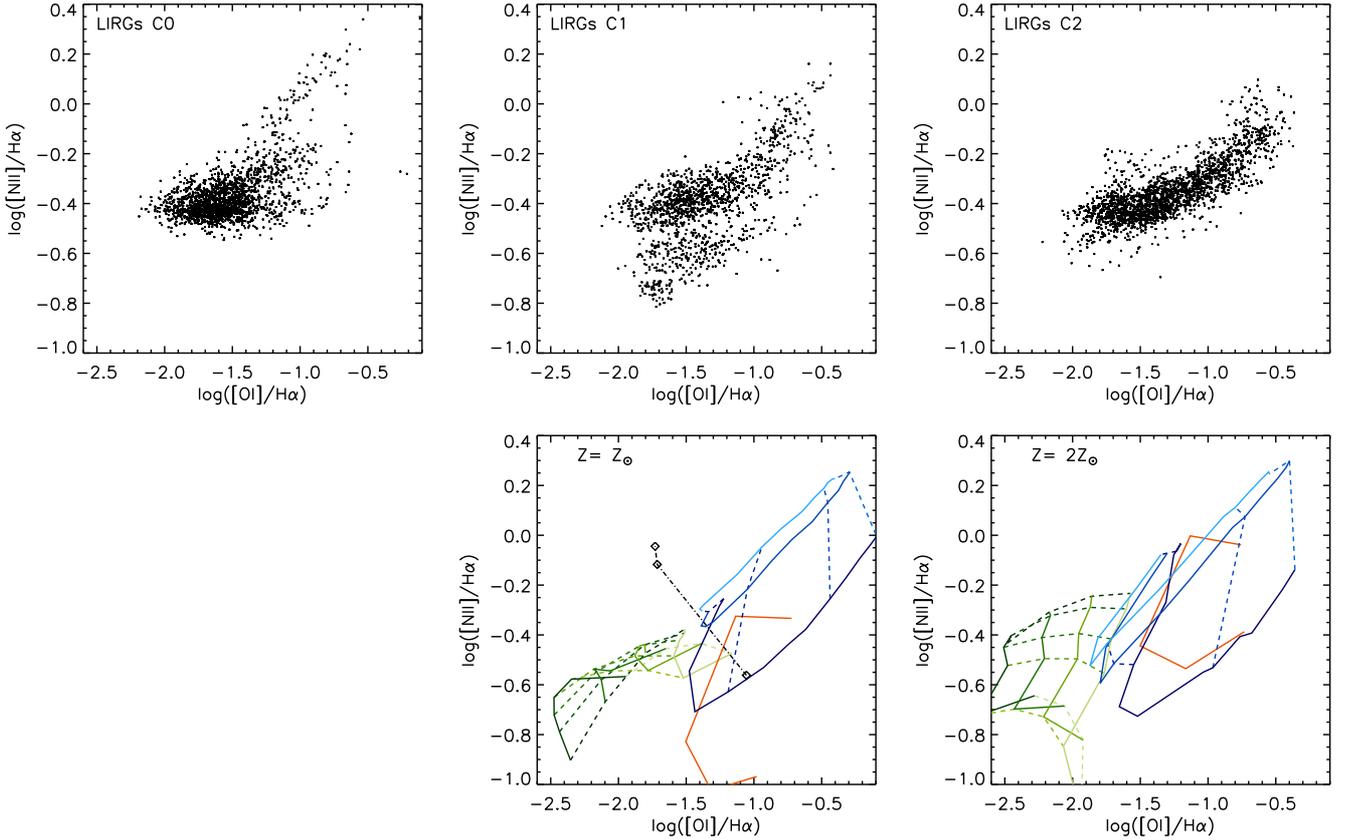}
   \caption[\oha\ vs. \nha.]
   {\oha\ vs. \nha\ diagrams. \emph{First row:} Line ratios for the
     different LIRGs 
     groups in our sample. \emph{Second row:} Theoretical models used
     for comparison where those in 
   the middle column have solar metallicity and those in the right twice solar
   metallicity.
The utilized color/symbol code is the same as in Fig. \ref{figs2havsn2gha}.
    \label{figo1havsn2gha}}
\end{figure*}

\subsection{Ionization mechanisms in the extra-nuclear ionized regions:
   young stars and shocks\label{n2gs2}} 

To investigate the nature of the ionization mechanisms present in the 
extended extra-nuclear regions, the \sha\
vs. \nha\ and  \oha{} vs. \nha{} diagnostic diagrams for (thousands)
of data points sampling these regions were compared with model
predictions for \ion{H}{ii} regions, shocks and AGNs.  
State-of-the-art models for evolving \ion{H}{ii} regions
\citep{dop06}, shocks \citep{all08}, TML \citep{sla93} and dusty AGNs
\citep{gro04} 
covering a wide range of physical parameters (ionization, density,
stellar ages, shock velocity, etc.) and metallicities were
considered. The results for the galaxies were separated according to their
interaction class, and set of models best representing the data are given
in Figs. \ref{figs2havsn2gha} and \ref{figo1havsn2gha} for the
\sha{} vs. \nha\ and  \oha{} vs. \nha{} diagnostic diagrams,
respectively. Galaxies with dubious 
  classification (see Table \ref{galaxias}) were excluded. The
data associated with each individual pointing have been attached as \emph{On-line
  Material}. 
 
From these diagrams it is clear that although the ionization properties
of most of the extended, extra-nuclear regions are consistent with
those of H\,\textsc{ii} regions, there is a trend with the interaction
class. As the interaction progresses (e.g. moving from class 0 to class
2), a larger fraction of extra-nuclear regions shows a clear shift towards an
increased excitation, as traced by the increment of the \sii\
and, in particular, \oi\  emission with respect 
to H$\alpha$ (see Figs. \ref{figs2havsn2gha} and \ref{figo1havsn2gha}). 

Although individual galaxies within an interaction class or regions
within galaxies can behave differently, a direct comparison of all
the available data (represented by thousands of points in the diagrams
of Figs. \ref{figs2havsn2gha} and \ref{figo1havsn2gha}) with the  
model predictions allow us to obtain some general conclusions about the
ionization mechanisms playing the main role in the extra-nuclear
regions of LIRGs.    

The first conclusion that can be drawn from a comparison of
  Figs. \ref{figs2havsn2gha} and \ref{figo1havsn2gha} is that TML
  are not playing a major role in the ionization of the extra-nuclear
  regions of this sample. Models can marginally reproduce the observed
  \nha\ vs. \sha, but clearly under-predict the \oha\ line ratio for a
  given \nha.

Instead, most of the regions are consistent with ionization by
young stellar populations with ages of 3~Myr or less and
metallicities twice solar. The measured H$\alpha$ emission line
equivalent widths in these regions are in the 40 to 300~\AA\ range
(Rodr\'i{\i}guez-Zaur\'in et al. in prep.), and therefore do not
correspond with the strong H$\alpha$ lines (equivalent widths of 1000  
to 2000~\AA) expected in very young ($\leq$ 3~Myr) stellar clusters
\citep[STARBURST99,][]{lei99}.
This apparent discrepancy is likely due to contamination by evolved
stellar populations \citep[see][]{alo10}. Because the physical scale of each resolution 
element (determined by a seeing of $\sim1\farcs2$) would be about 480
pc, the measured  equivalent 
widths are likely the result of different stellar populations where the  
old stars present in these regions contribute to the optical continuum,
but not to the emission lines. Moreover, typical sizes
for giant \ion{H}{ii} regions would be $\lsim$400~pc
\citep[e.g.][]{ken84,alo02}. Because our typical resolution is slightly higher,
a single spaxel can sample both pure DIG and regions of very young star
formation. This  implies that measured line ratios, specially the
\nha\ one, are higher than those for pure \ion{H}{ii} regions and thus a
comparison with models suggests too young stellar populations. 

For regions showing a $\log$(\sha)$\gsim-0.6$, the
model predictions considering ionization mechanisms different
 from the TML's ones
overlap and therefore no firm conclusions can be made based on the
\sha\ vs. \nha\ alone (see Fig. \ref{figs2havsn2gha}). However,
much of this overlap disappears when the \oha\
ratio is considered. In particular regions showing
 $\log$(\oha)$\geq -1.6$ are best explained by the
presence of shocks with velocities of less than 200 km s$^{-1}$ and
metallicities 1-2 solar (see Fig. \ref{figo1havsn2gha}).  

We note that a detailed comparison of our data with the
  models in Figs. \ref{figs2havsn2gha} and \ref{figo1havsn2gha} may
  lead to some inconsistencies.  For instance, the \sha
- \nha\ diagram presents less data points in the
area of shocks of
$v_s\sim200-300$~km~s$^{-1}$ than the \oha - \nha\ diagram. 
These disagreements may be due to the intrinsic difficulty  in
modeling some emission \citep[e.g. \textsc{[O\,i]}$\lambda$6300
  line][]{dop95} as well  as  possible observational effects. 
The areas sampled by a given spaxel include both star-forming
regions and diffuse ionized gas, which affect the line
ratios differently. In any case, these figures should be read looking for general
behaviors rather than for detailed comparisons. 

Taking this into account, we considered the \oha - \nha\
diagram  as  our main indicator to establish the most important
ionization mechanism for 
individual galaxies in Table \ref{mecanismo}. In some cases this was
complemented with other indicators, especially for the galaxies with
no \oi \ detections. 

Elevated line ratios can in principle be explained by both
  shocks and AGN. This is evident from the models presented in Figs.
  \ref{figs2havsn2gha} and \ref{figo1havsn2gha}, as well as from
  integrated data from the literature \citep[][]{kew06}. 
Indeed, four of the 32 galaxies are classified as Seyfert (the northern
member of \object{IRAS~F07027-6011}, \object{IRAS~F13229-2934}, and
the western member of \object{IRAS~14544-4255} and
\object{IRAS~21453$-$3511}, see footnotes in Table \ref{mecanismo}).
The northern member of \object{IRAS~F07027-6011} 
should not pose a problem, because most of its line ratios are typical
of young stars. However, as stated in Table \ref{mecanismo}, some
areas with elevated line ratios could be caused by the nuclear AGN. The galaxy
\object{IRAS~13229$-$3934} contains a central AGN
which causes an ionization cone in the north-south direction
  \citep{bed09}. However, our measured line ratios are associated to
  wide areas in the \emph{inter-arms} zone, not coincident with the
  direction of the ionization cone, and thus difficult to be explained
  by the central AGN. Finally, for the western member
  \object{IRAS~14544-4255} and \object{IRAS~21453$-$3511}, it is not
  possible to discern with the present information whether the
  ionization in the external areas is caused by shocks or the central
  AGN. However, an indirect argument allows us to favor shocks against
  AGN as the mechanisms responsible of the observed line ratios. As 
was shown for  Arp~299, a very nearby LIRG, the
extra-nuclear area with an 
excitation caused by the AGN (as seen by the \ohb\ line ratio) is very
small and restricted to specific directions defined by the ionization
cone \citep[see detailed analysis for this system in][]{gar06}. Also,
nuclear regions, where one could expect the largest influence of an AGN,
were removed from our analysis (see Sect. \ref{secdata}). 

It is worth mentioning that the metallicities (solar to twice solar) of
the models (\ion{H}{ii} and shocks) that best fit the range of data
are underabundant by a factor of $\sim1.5-2.0$  with respect to what
is expected from the mass-metallicity relation for galaxies
\citep{tre04}. This relation predicts metallicities of $Z \sim 1.9 - 2.8$~Z$_\odot$\footnote{We have
  employed  $12 + \log ($O/H$)_\odot = 
  8.66$, from \citet{asp04}.} for  galaxies with masses as
those expected in typical LIRGs \citep[$\sim 10^{10} -
  10^{12}$~M$_\odot$, e.g.,][]{hin06,vai08b,vai08a}. Similar findings for the nuclear region
of these kinds of galaxies have been reported by \citet{rup08}.

\begin{table*}[h!]
\centering
            
\caption[]{Ionization in the extra-nuclear ionized regions of LIRGs.
     \label{mecanismo}}
\scriptsize
 \begin{tabular}{cllp{8cm}}
       \hline
       \hline
            \noalign{\smallskip}
  Galaxy       &  Mechanism$^{(a)}$  & $r$ $^{(b)}$ & Comment \\
 (IRAS number) &   &  &    \\
\hline
\multicolumn{4}{c}{Class 0}\\
\hline
F06295$-$1735    &  Young stars  & 0.22 & \\
F06592$-$6313    &  \ldots       & 0.42 & No \oi\ nor \sii\ data. $r$ based on \nha.\\
F07027$-$6011~N$^{(c)}$  &  Young stars and shocks (100-200~km~s$^{-1}$) & 0.44 &  Part of the shocks may be associated to the nucleus according \oha \ map.\\
F07027$-$6011~S  &  Young stars  & 0.21 &  $r$ based on
       \sha. $r$ based on \oha\ not reliable.\\
F07160$-$6215    &  Shocks (100-300~km~s$^{-1}$) & 0.31 & $r$ based on \sha. No \oi\ data.\\ 
F10015$-$0614    &  Young stars & 0.15 & \\
F10409$-$4556    &  \ldots       & 0.37 & $r$ based on \nha.  No \oi\  nor \sii\ data.\\
F10567$-$4310    &  Young stars  & 0.23 & $r$ based on \sha. No \oi\ data.\\
F11254$-$4120    &  Young stars  & 0.71  & $r$ based on \nha.  $r$ based on \sha\  not reliable. No \oi\ data. Possible shocks associated to the bar according to the \sha \ map. \\
F11506$-$3851    &  Young stars  & 0.19  & Some small areas better explained with low velocity shocks.\\
F12115$-$4656    &  Young stars  & -0.24 & \\
F13229$-$2934$^{(d)}$     & Shocks  (100-300~km~s$^{-1}$)  & 0.47 & Some areas explained by stars, mostly associated with the arms. Part of the high excitation may be associated to the nucleus according to the \oha \ map. \\ 
F22132$-$3705    &  Young stars  & 0.10 & Small areas better explained with low velocity ($<$100~km~s$^{-1}$) shocks.\\
\hline
\multicolumn{4}{c}{Class 1}\\
\hline
F01159$-$4443    & Low velocity  ($<$100~km~s$^{-1}$) shocks & 0.23 & Some regions explained by stars.\\ 
ESO 297-G011     & Young stars   & -0.12 & Some small areas explained by shocks. $r$ based on \sha. $r$ based on \oha\ not reliable.\\
ESO 297-G012     & Young stars and shocks & -0.56 &  $r$
       based on \nha. $r$ based on \oha\ or \sha\ not reliable.\\
F06076$-$2139    & \ldots        & 0.75 & $r$ based on \nha \ data. No \oi\ nor \sii\  data.\\
F06259$-$4708 P1 & Young stars   & 0.74 & Low velocity (100-150~km~s$^{-1}$) shocks also present.\\
F06259$-$4708 P2 &  Shocks (100-150~km~s$^{-1}$) and young stars & 0.31 &  \\
08424$-$3130     &  Stars and shocks ($\sim100$~km~s$^{-1}$)  & 0.61 & $r$ based on \sha. $r$ based on \oha\  not reliable. Northern galaxy mainly shocks; southern one mainly stars. \\ 
F08520$-$6850    & Stars and shocks ($\sim100$~km~s$^{-1}$)   & 0.40 &  \\
IC563            & Young stars   &  0.22 & Low velocity (100-150~km~s$^{-1}$) shocks could be also present.\\
IC564 N          & Young stars   &  0.44 & Low velocity (100-150~km~s$^{-1}$) shocks also present. $r$ based on \nha. No \oi\ data. $r$ based on \sha\  not reliable.\\
IC564 S          & Young stars   &  0.34 & $r$ based on \nha.  No \oi\ data. $r$ based on \sha\  not reliable.\\
12042$-$3140     & Young stars and shocks (100-150 ~km~s$^{-1}$) &  0.81 & Northern galaxy mainly stars, Southern one shocks. \\
12596$-$1529     & Young stars and low velocity shocks & 0.47   & $r$
based on \sha\ data.  No \oi\ data. $r$ based on \sha\ not reliable.\\
F14544$-$4255 E  & Young stars   &  0.67 & $r$ based on \sha. No \oi\ data.\\ 
F14544$-$4255 W$^{(d)}$   & High velocity (100-300~km~s$^{-1}$) shocks & 0.67 & Shocks may be associated to the nucleus according to the \oha\ map.\\ 
18093$-$5744 N   & Young stars   & 0.09 & Low velocity shocks also possible.\\
18093$-$5744 C   & Stars older than 3~Myr and/or with $Z <Z_\odot$ & -0.65 &$r$ based on \sha. $r$ based on \oha\ not reliable.\\
18093$-$5744 S   & Young stars   & 0.27 &     \\
\hline
\multicolumn{4}{c}{Class 2}\\
\hline
F04315$-$0840    &  Shocks (100-200~km~s$^{-1}$) and young stars & 0.63 & Young stars in the inner region and arms/tidal tails. Shocks at 2-3 kpc from the nucleus.  \\ 
08355$-$4944     &  Young stars & 0.14 & Low velocity shocks also present at $\sim$ 2 kpc from the nucleus. \\
F10038$-$3338    &  Shocks (100-200~km~s$^{-1}$) & 0.50 & Shocks may be associated to the nucleus. \\  
10257$-$4338     &  Shocks (100-150~km~s$^{-1}$) and stars &  0.63 &  \\
12116$-$5615     &  Young stars and low velocity shocks & 0.57 & \\
13001$-$2339     &  Shocks (100-150~km~s$^{-1}$) & 0.48 & All data with \oha$>-1.2$. Part of the shocks may be associated to the nucleus  \\
17138$-$1017     &  Young stars and low velocity shocks & -0.23 & No \oi data; $r$ according to \sha \ data \\
21453$-$3511$^{(d)}$     &  High velocity (100-300~km~s$^{-1}$) shocks and/or AGN. & 0.76 & Young stars (spiral arms). Shocks may be associated to the nucleus according \oha\ map.\\ 
\hline
            \hline
 \end{tabular}
\begin{flushleft}
$^{(a)}$ At least otherwise indicated, based on the comparison of
  the data with several ionization models in the \oha - \nha \
  diagnostic diagram.  For the present analysis solar metallicity has
  been considered as baseline. If higher metallicity
  (e.g. Z=2~Z$_\odot$) is considered, a higher presence of shocks with
  respect to stars would have been obtained (see
  Figure~\ref{figo1havsn2gha}). \\  
$^{(b)}$ Pearson coefficient inferred from the \oha - $\sigma$
  relation (see text). For objects where \oi\ was not detected or the
  $r$ coefficient did not reach a confidence level of 90\%, $r$
  was obtained from the \sha \ or \nha, as indicated.\\   
$^{(c)}$ Classified as Seyfert \citep{kew01a}.\\
$^{(d)}$ Classified as Seyfert \citep{cor03}.\\
\end{flushleft}

\end{table*}

In summary, the ionization of the extended, extra-nuclear regions in
isolated galaxies (class 0), is mostly explained as due to young stars
like in \ion{H}{ii} regions. Systems showing some degree
of interaction (class 1 and 2)  
present a large and increasing fraction of regions which are better 
explained by shocks. This is particularly evident when using the  
the \oha{} vs. \nha{} diagnostic diagram.  
In particular cases (e.g. IRAS~21453$-$3511), an AGN also
could explain the observed high line ratios. 
In order to explore the precise relevance of a putative AGN, other additional
observables than those utilized here (e.g. flux in the
\textsc{[O\,iii]}$\lambda$5007 emission line) would be needed.
Further evidence for the presence and relevance of shocks as ionizing sources 
can be found through the gas velocity dispersions and their
correlations with the emission line ratios. This is presented in the
next section. 

\subsection{The relation between excitation and gas velocity
  dispersion: further evidence for the importance of shocks}  

\begin{figure*}[!ht]
   \centering
\includegraphics[angle=0,width=0.99\textwidth, clip=,bbllx=60, bblly=255,
bburx=508, bbury=658]{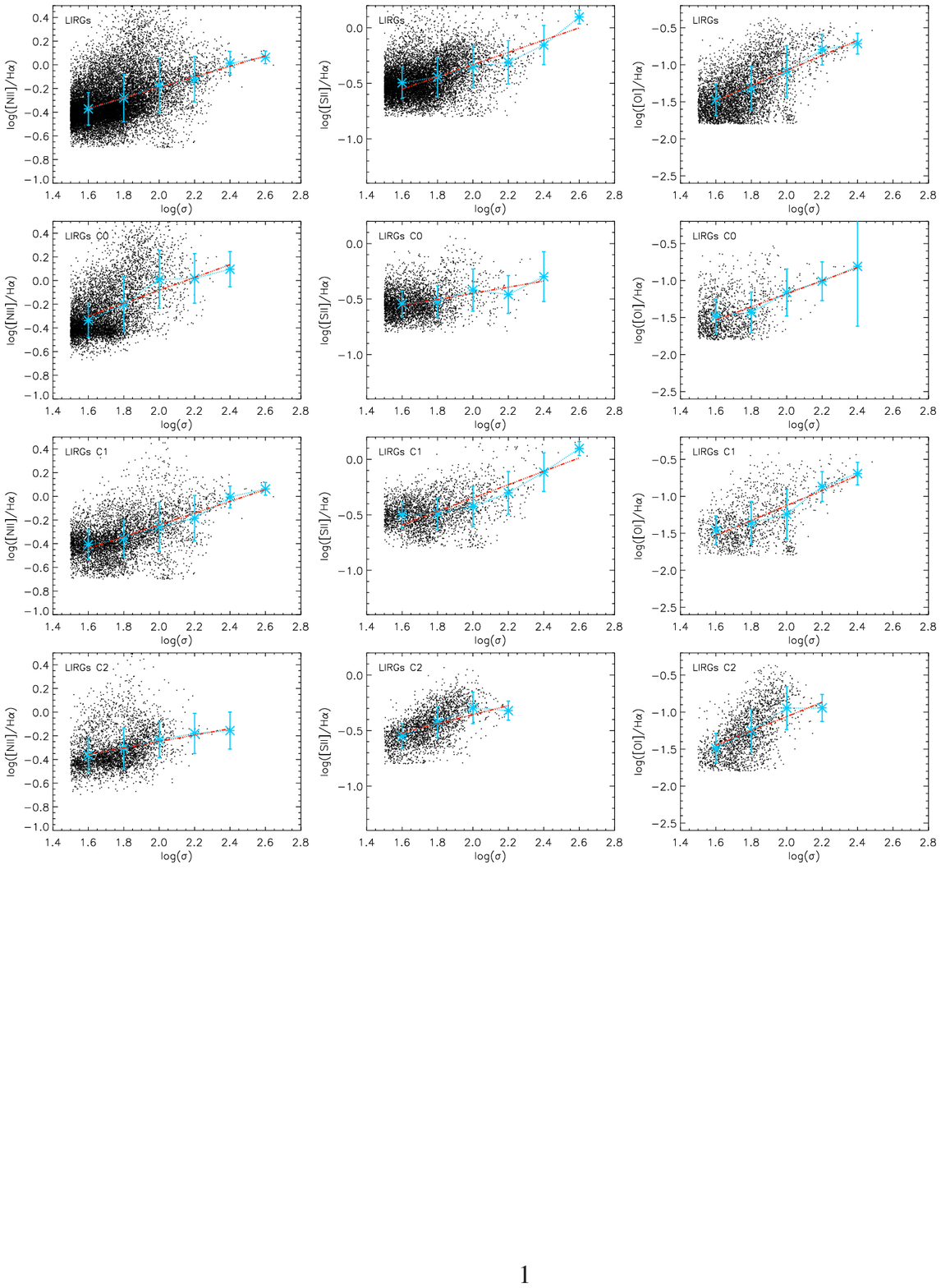}
   \caption[Velocity dispersion vs. line ratios.]
   {Relation between the velocity dispersion and the
   [N\,\textsc{ii}]$\lambda$6584/H$\alpha$ (left),
   [S\,\textsc{ii}]$\lambda\lambda$6717,6731/H$\alpha$ (middle), and
   [O\,\textsc{i}]$\lambda$6300/H$\alpha$ (right). Mean and
     standard deviation of each 0.2~dex bin in the velocity dispersion
     are shown with blue crosses and error bars respectively.
The red dash-3 dotted line in each individual graphic represent the
  one-degree polynomial fit to these values (see Table
  \ref{pearsonvimos} for specific values).
   We refer the reader to the
   \emph{On-line only} version for the equivalent plots associated with
   each individual system.
    \label{cocivsdisp}}
\end{figure*}

   \begin{table*}
\centering
      \caption[]{Linear Pearson correlation coefficients $r$ and
      1-degree polynomial fits for the different
      morphological groups considered here.
      \label{pearsonvimos}} 
              \begin{tabular}{ccccccccccccc}
            \hline
            \noalign{\smallskip}
Group & 
\multicolumn{2}{c}{[N\,\textsc{ii}]$\lambda$6584/H$\alpha$} &
\multicolumn{2}{c}{[S\,\textsc{ii}]$\lambda$6717,6731/H$\alpha$} &
\multicolumn{2}{c}{[O\,\textsc{i}]$\lambda$6300/H$\alpha$}\\
& r & A+Bx &  r & A+Bx & r & A+Bx \\
\hline
All     & 0.43 & -1.08 + 0.44 x &  0.33 & -1.44 + 0.55 x & 0.54 & -3.12 + 1.02 x\\
Class 0 & 0.54 & -1.17 + 0.54 x &  0.23 & -1.00 + 0.28 x & 0.30 & -2.95 + 0.88 x\\
Class 1 & 0.43 & -1.23 + 0.50 x &  0.33 & -1.55 + 0.60 x & 0.48 & -3.14 + 1.01 x\\
Class 2 & 0.31 & -0.79 + 0.27 x &  0.60 & -1.18 + 0.69 x & 0.61 & -3.00 + 0.97 x\\
            \hline
         \end{tabular}
  \end{table*}

The presence and relevance of shocks in (U)LIRGs has already been
suggested by a positive relation between the ionized
gas velocity dispersion and its ionization degree as traced by the
\sha\ ratio in a sample of about 30 galaxies
\citep{arm89,dop95,vei95}. These studies were based on long-slit 
observations and were therefore dominated by the contribution from
the high-surface brightness nuclear regions in a large number of
cases. Also, the slit was positioned along a given orientation 
and therefore the results of these studies do not necessarily
represent the excitation and kinematics of the ionized gas in the
extra-nuclear extended (several kpc) regions. Instead, for the
detailed study of the excitation conditions in the extended regions it
is more appropriate to use the  two-dimensional information provided
by IFS data once the nuclear regions 
are removed, as already discussed for a small sample of ULIRGs
\citep{mon06}. We performed a similar study with the present
  sample of LIRGs, limiting the analysis to the data with \nha, \sha,
  and \oha\  higher than $-0.7$, $-0.8$, $-1.8$, respectively
  (i.e. line ratios in the range expected for shocks according to the
  models discussed in the previous section). In particular, Fig.
  \ref{cocivsdisp} presents the relation between the excitation
degree (here represented by our three line ratios) and the gas
velocity dispersion for the entire sample (upper panels) and 
according interaction class (three lower panels).
The relations for
each individual pointing  have been appended in the \emph{On-line only}
version (see Fig. \ref{lirgs_indi}). 

Several conclusions are already evident from these results. Firstly, the
ionized gas in the extra-nuclear regions of LIRGs has typically velocity
dispersions between 32 km s$^{-1}$ (lower limit given by our  
spectral resolution) and 125 km s$^{-1}$, with very few regions having
velocities above this value, independently of the morphology
  and infrared luminosity.

Secondly, the velocity dispersion of the
  ionized gas  is larger in galaxies with some degree of
  interaction, i.e. class C1 and C2. While galaxies identified as
  isolated have a median velocity dispersion of 37 km~s$^{-1}$, class C1 and
  C2 have values of 46 km~s$^{-1}$ and 51 km~s$^{-1}$, respectively.
  Moreover, the 
  velocity dispersions of class 0 galaxies (i.e. isolated) tend to be
  concentrated in the low velocity range, while classes 1 and 2 have a
  relevant fraction of regions with high velocities.
This means that  only 5\% of the data points for
class 0 galaxies present velocity dispersions larger than 80 km s$^{-1}$,
while 28\% and 16\% of those for classes 1 and 2
respectively do when considering, for example, the \oha\ line ratio.
These values are much higher than the median velocity dispersions
measured  in the extranuclear ionized regions of normal galaxies with
velocities in the 20 to 30 km~s$^{-1}$ range \citep{epi10}. This is a
clear indication that the  ionized interstellar medium in LIRGs in
general, and even more in interacting LIRGs, is dynamically hotter
than the quiescent  ISM of normal galaxies due to the
strong shocks produced by the tidal forces and by stellar winds in the
powerful nuclear starbursts.

Thirdly, there is a clear correlation between the
excitation degree and the velocity dispersion in interacting galaxies
(class 1) and mergers (class 2), while the evidence of
  correlation in isolated galaxies (class 0) is poor and restricted to
  the relation involving the \nha\ line ratio.


To obtain a more quantitative analysis of the degree of the correlation, the 
linear Pearson correlation coefficient, $r$, is used. This coefficient 
quantifies the degree of correlation between two given quantities
that are assumed to follow a linear relation. It varies from $-1$ to $+1$,
where $+1(-1)$ means a perfect correlation(anti-correlation) and 0
means no-correlation. The Pearson coefficient for the
$\log$(\sha) - $\log (\sigma)$ in the  \citet{arm89} sample \citep[see
also Fig. 8 in][]{dop95}
has a value of $r=0.4$. Hereafter we consider
that a positive relation exists between the excitation conditions and
velocity dispersion in the ionized gas only if $r$ is higher than 0.4. The
computed $r$ coefficient for the different groups as well as the polynomial 
coefficients obtained from  a least-square one-degree polynomial fit to
the mean of the data in 0.2~dex bins in the velocity
  dispersion appear in Table \ref{pearsonvimos}. For the 
  individual pointings, the direct fits to the data are
  included in the lower right 
  corner of the corresponding panel in Fig. \ref{lirgs_indi} in
  the \emph{On-line only} version. 

\begin{figure}[!ht]
   \centering
\includegraphics[angle=0,scale=.8, clip=,bbllx=20, bblly=5,
bburx=280, bbury=435]{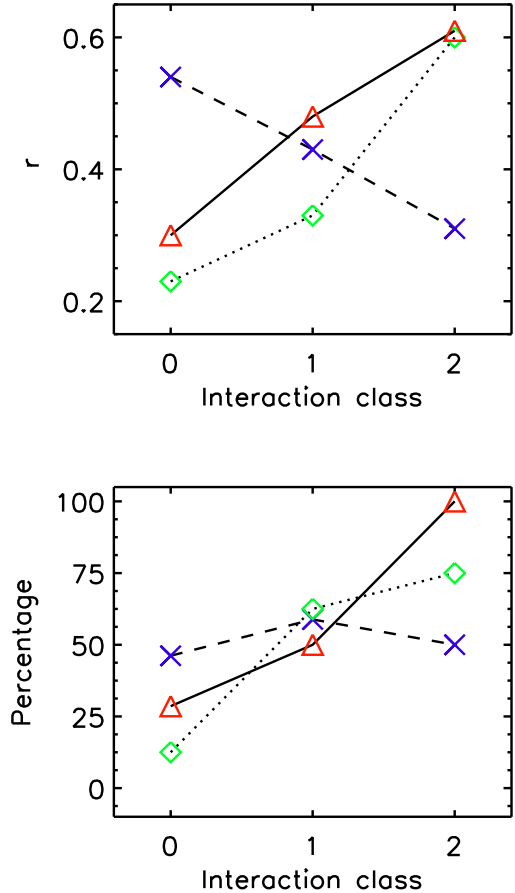}
   \caption[Interaction class vs. mean $r$ coefficient ratio and percentage.]
   {Variation of the $r$ coefficient (up) and percentage of
   galaxies with a confidence level higher than 90\% showing
   correlation (down) with the interaction types 
   (0=isolated, 1=interacting pairs, 2=merger remnants) for the three
   emission line ratios. We used the same line and color code as in 
   Fig.~\ref{porcenhistogramas} to distinguish among the
   emission line ratios. \label{intervsr}} 
\end{figure}

The confidence level of the different relations was estimated by
means of the t-Student test.  
All relations corresponding to the
  different groups present a confidence level higher than 99.9\%. For
  individual pointings, the \nha-$\sigma$ show a larger degree of
  confidence than those for the \sha-$\sigma$ and \oha-$\sigma$
  relations, because the S/N limits the number of points.
  All relations involving the \nha\ line
  ratio except the southern member of
  \object{IRAS~18093$-$5744} present a level 
  of confidence higher than 95\%. For the other line ratios the
  level of confidence is in general lower.
  For the \sha\ line ratio, 11
  out of 36 pointings (i.e. 30\%) did not reach a 90\% of confidence
  level. These are \object{IRAS~F06295$-$1735},  
  \object{IRAS~F11255$-$4120},
  \object{IRAS~F11506$-$3851}, \object{IRAS~F01159$-$4443},
  \object{ESO~297$-$G012},  \object{IC~563}, both pointings for
  \object{IC~564},  
  the northern and central member of \object{IRAS~18093$-$5744} and 
  \object{IRAS~12596$-$1529}.
In the case of the relation involving
  the \oha\  line ratio 22\%  of the pointings (i.e. 6 out of 27) did
  not reach a 90\% confidence level. They are the southern member of
  \object{IRAS~F07027$-$6011}, \object{ESO~297$-$G011},
  \object{ESO~297$-$G012}, \object{IRAS~08424$-$3130}, the central member of 
  \object{IRAS~18093$-$5744} and \object{IRAS~12596$-$1529}.

The variation of the Pearson coefficient ($r$) with the interaction class
 is summarized in the upper part of Fig. \ref{intervsr}.
The percentage of galaxies in each group showing a correlation
coefficient $r>0.4$ and a level of confidence higher than 90\% 
is summarized in the lower part of Fig. \ref{intervsr}.
The quantitative analysis emphasizes the main differences according to the 
interaction class already mentioned above.  
The degree of correlation between the excitation properties and the
velocity dispersion of the ionized gas when the \sha\ and
  \oha\ line ratios are considered increases with the 
  degree of interaction.
In particular, the \oha\ ratio is the best tracer of
shocks as already shown in the previous section. When using this
ratio, the Pearson correlation coefficient for the class 
0 galaxies is $r= 0.30$, indicating no correlation. This result
supports the idea that shocks are not playing an important role in the
ionization of the external areas of this group of LIRGs and agrees well with
our previous findings  that young stars are the main cause for
the ionization. On the other hand, the mean $r$ 
values for classes 1 and 2 (0.48 and 0.61, respectively) are
well above 
the 0.4 criterion, which we considered necessary for a good positive relation. This
quantitative result clearly supports a direct cause-effect relation
between the dynamical status of the gas, i.e. turbulence and shocks  
traced by the velocity dispersion, and its excitation conditions. 

In summary, there are two clear differences between LIRGs classified
as class 0 (i.e. isolated), and those classified as class 1 (mostly pairs), and
class 2 (mergers). The ionized gas in classes 1 and 2 is
characterized by covering a wide range in velocity dispersion with an
extension towards higher values than class 0 galaxies. Moreover, the
dynamical status of the gas, turbulence and shocks, plays an important
role in the excitation of the gas mainly in  LIRGs classified as
interacting pairs or evolved mergers.   

\begin{figure*}[!ht]
   \centering
\includegraphics[angle=0,width=0.99\textwidth, clip=,bbllx=60, bblly=250,
bburx=508, bbury=660]{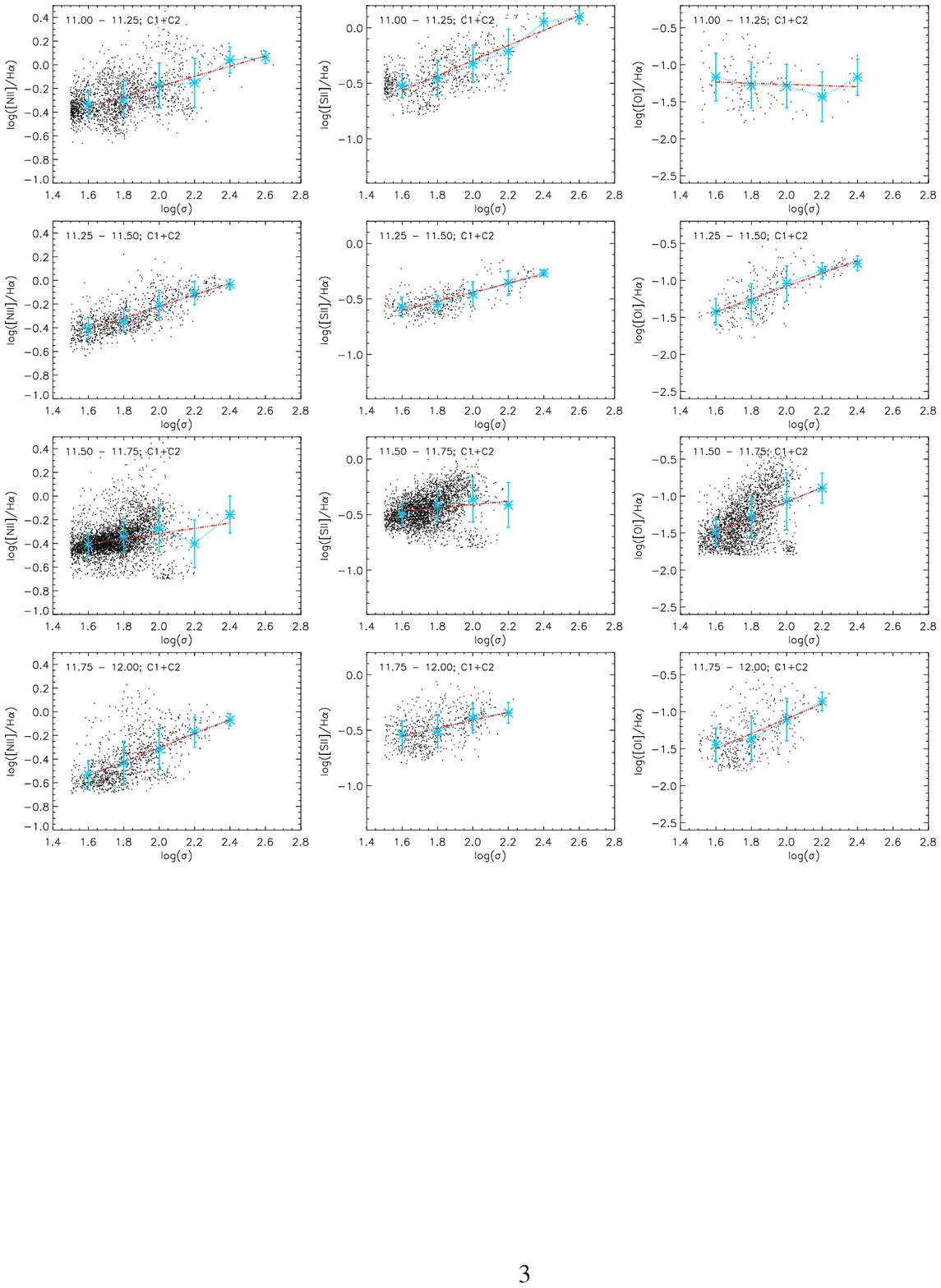}
   \caption[Velocity dispersion vs. line ratios in the different
   luminosity bins.]
   {Same as Fig. \ref{cocivsdisp} but for class 1 and 2 systems
   grouped in four luminosity bins.  
    \label{cocivsdisp_lumi}}
\end{figure*}

   \begin{table*}
\centering
      \caption[]{Linear Pearson correlation coefficients $r$ and
      linear fits for for the class 1 and 2 LIRGs distributed
        in luminosity groups spanning the entire luminosity
        range. \label{pearsonvimos_lumi}} 
              \begin{tabular}{ccccccccccccc}
            \hline
            \noalign{\smallskip}
Luminosity Range & 
\multicolumn{2}{c}{[N\,\textsc{ii}]$\lambda$6584/H$\alpha$} &
\multicolumn{2}{c}{[S\,\textsc{ii}]$\lambda$6717,6731/H$\alpha$} &
\multicolumn{2}{c}{[O\,\textsc{i}]$\lambda$6300/H$\alpha$}\\
 ($\log(L/L_\odot)$)& r & A+Bx &  r & A+Bx & r & A+Bx \\
\hline
11.00-11.25 & 0.49 & -0.97 + 0.39 x &  0.63 & -1.39 + 0.54 x & -0.05 & -1.12 - 0.09 x\\
11.25-11.50 & 0.72 & -1.24 + 0.51 x &  0.63 & -1.19 + 0.37 x & 0.68 & -3.09 + 1.01 x\\
11.50-11.75 & 0.28 & -0.77 + 0.22 x &  0.24 & -0.81 + 0.20 x & 0.35 & -2.47 + 0.61 x\\
11.75-12.00 & 0.59 & -1.58 + 0.64 x &  0.43 & -1.21 + 0.41 x & 0.46 &
-3.01 + 0.95 x \\
            \hline
         \end{tabular}
  \end{table*}

\subsection{Tidal forces as the origin of shocks in the extra-nuclear regions?} 

It was shown above that the importance of shocks in
large extended regions increases with the interaction class. What is
the origin of these shocks? Are the shocks caused by the tidal forces
due to the interaction process itself? 
Or are the shocks produced in stellar superwinds associated with the intense  
starbursts generated in the nuclear regions?  
 
To answer these questions, we investigated how the relation between the
excitation conditions and gas velocity dispersion varies with the
total infrared luminosity of the system.
In the sample considered here, only four objects were
  classified as Seyfert - and thus harboring an AGN- according to their nuclear
spectra \citep{kew01b,cor03} (see Table \ref{mecanismo}).
Thus the total infrared luminosity is considered in general a direct tracer 
of the intensity of the star formation, because the infrared luminosity in
these galaxies scales linearly with the star-formation rate
\citep[e.g.][]{ken98}. 
Figure \ref{cocivsdisp_lumi}  presents  similar relations to those
in Fig. \ref{cocivsdisp} for those classes with some degree of
interaction (class 1 and 2) but this time 
binned in four luminosity ranges covering the entire luminosity range
of LIRGs (i.e. from 10$^{11}$ to 10$^{12}$ L$_{\odot}$).

Contrary to what
happens with the interaction class \emph{there is no evidence for a
  dependence of the correlation coefficient with the
  luminosity bin}, independently of the utilized line ratio.
Assuming star-formation dominates the energy output in these galaxies
(see above), the range in infrared luminosity covered by the sample
represents a change of a factor ten in the star-formation rate
\citep[see][for the specific relation]{ken84}. Therefore the
radiative and mechanical energy released in the surrounding ISM due to 
supernovae explosions and stellar winds  produced in young massive
stars would increase linearly with the rate of star formation
\citep[see e.g.][]{col91}. These linear relations in the mass,
momentum and energy deposition have already been measured in the cool,
neutral gas traced by the \ion{Na}{i} line \citep{rup05a}. However,
these studies do show some evidence for a flattening of these
relationships for star-formation rates above 10 M$_\odot$~yr$^{-1}$
(i.e. LIRG and ULIRG range).  It is unclear whether these results
would apply to the ionized gas traced by the \ha\ line.  The momentum
and energy release in the cool, neutral gas is usually a small
fraction of the energy in the warm ionized and hot X-ray emitting
gas. Moreover, the nonlinear relation between the outflow velocity of
the cool gas and the SFR \citep[$\propto$SFR$^{0.35}$,][]{mar05} could
indicate a saturation in the mechanical energy liberated into the more
dense neutral gas, but not necessarily in the other phases of the
ISM. In addition, the velocity gradients measured in several outflows
detected in ULIRGs are inconsistent with the expected gaseous radial
flows produced by nuclear (size of 200-300 pc) starbursts
\citep{mar06}. These gradients could indeed still be consistent with
more extended starbursts on scales of kpc, or shocks
generated by tidal forces during the interaction process \citep{mar06}. 
Our H$\alpha$ IFS of (U)LIRGs indicates that the H$\alpha$ emission is more concentrated than that of the stellar continuum. In particular, the fraction of H$\alpha$ emission within the central 2 kpc is higher than that of the continuum for  about 80\% of the cases. However,  $\sim$60\% of the objects have more than half of their H$\alpha$ emission outside the central 2 kpc  (see Rodr\'{\i}guez-Zaur\'{\i}n et al. in prep. for details).

Therefore, if the detected evidence
of shocks were due to star formation, it would be reasonable to expect a more
turbulent gas with stronger outflows and shocks. This could be traced
by higher excitation conditions and velocity dispersions in the
ionized gas as well as higher correlation degrees. No evidence
for any of this is observed in the present data that sample large
extra-nuclear regions of several hundreds of pc to several kpc
  in size, outside the circumnuclear regions.   

The comparison between Figs. \ref{cocivsdisp}  and
\ref{cocivsdisp_lumi}  indicates that the presence and relevance 
of shocks are more strongly correlated with the interaction/merging
class of a system than with its star formation activity.  
Indeed, our sample of class 0 galaxies shows on average a level of
star formation activity (i.e. infrared luminosity) similar to or
slightly lower than that of classes 1 and 2 ($\log (L/L_\odot)= 11.44$ for
class 0 against 11.51 and 11.60 for classes 1 and 2, respectively). 
Note that this statement is valid for our particular sample, which was
selected in order to cover all the interaction types and luminosity
ranges in a more or less uniform way. However, it does not apply to
complete samples of LIRGs, because they show a much higher percentage of
interaction/merging systems at higher luminosities \citep[e.g.][]{san04}.
These results point to tidal forces associated with the interaction/merging
process as the origin for the shock ionization in the extended,
extra-nuclear regions.   
A similar result was found by MAC06, who concluded that the more likely
explanation for  shocks in five out of the six ULIRGs
studied there were tidally induced large scale gas
flows caused by the merging process. 

Detailed studies of the nearest ULIRG \object{Arp~220}
\citep{col03,mcd03}  have also suggested that large
extended regions in this system arise purely from merger dynamics and
collisional shock heating of the gas.  At the same time, footprints
associated with starburst superwinds have also been
detected \citep[][]{hec90,arr01}.  
In Arp~220, the ionized gas plumes that could be associated with
superwinds generated in  the nuclear starburst form an elongated 
structure up to a distance of about 2~kpc from the nucleus. However, for
the present sample of LIRGs, which show a significant lower star-forming
activity based on their L$_{IR}$ but similar dynamical
mass \citep{hin06,vai08a,vai08b} and hence similar escape velocity, a
smaller area of influence of the stellar superwinds
should be expected.  

Thus our results indicate that the tidal forces during the
interaction process are the mechanism producing the ionizing shocks in
the extended extra-nuclear regions in LIRGs. This is still compatible with 
the existence  of SGWs produced in nuclear starbursts at distances closer to the nucleus (i.e. radius of 1 to 2 kpc), or AGN
ionizing cones along particular orientations \citep[e.g. Arp~299,][]{gar06}.

\subsection{Interacting LIRGs and ULIRGs. Towards a common
  $\log$(\oha) - $\log (\sigma)$ relation?}  

\begin{figure*}[!ht]
   \centering
\includegraphics[angle=0,width=0.99\textwidth, clip=,bbllx=60, bblly=350,
bburx=508, bbury=560]{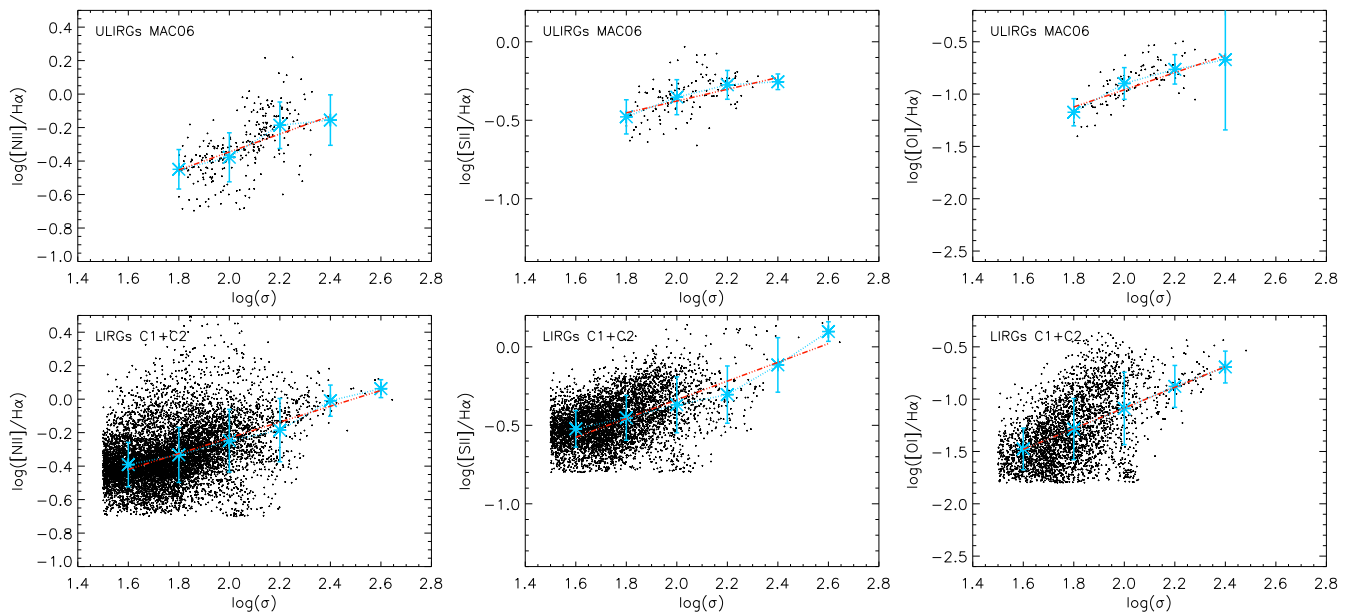}
   \caption[Velocity dispersion vs. line ratios for INTEGRAL data.]
   {\emph{Upper row:} Same as Fig. \ref{cocivsdisp}, but for all 
   systems in MAC06 but 
   \object{IRAS~17208$-$0014}. This is similar to the first
   row of Fig. 4 of MAC06 but we plotted the same ranges in the
   axes as in Fig. \ref{cocivsdisp} for a better comparison with the
   results in the present paper. \emph{Lower row:} Same as Fig.
   \ref{cocivsdisp} for the combined class 1 + class 2
   group.
    \label{cocivsdisp_integral}}
\end{figure*}

In previous sections we showed that our IFS data indicate that
shocks produced by the tidal forces in interacting and
  merging LIRGs play
a relevant role in the excitation of the extended ionized regions,
without any clear relation with the intensity of the star
formation. Because ULIRGs are the extreme cases of interactions and
mergers, one may wonder how the LIRGs situation compares with that
for the ULIRGs. For that purpose, we compared our results  with those
presented in MAC06.  
The relatively small sample of MAC06 (i.e. six ULIRGs systems, nine
galaxies) was made out of three class 1 systems
(\object{IRAS~08572+3915}, \object{IRAS~12112+0305},
\object{IRAS~14348$-$1447}) and three class 2 systems
(\object{IRAS~15206+3342}, \object{IRAS~15250+3609},
\object{IRAS~17208$-$0014}). We found that in all the systems but
\object{IRAS~17208$-$0014} the extra-nuclear extended regions  were 
well explained by ionization due to fast shocks with velocities of
$150-500$~km~s$^{-1}$. 
Because none of the systems in the MAC06's ULIRGs sample is
  classified as isolated (class 0), we present in Fig.
  \ref{cocivsdisp_integral} (second row) the line ratio vs. velocity 
dispersion only for LIRGS of classes 1 and 2.

Considering the best shock tracer, the \oha\ line ratio, the similarity of the $\log$(\oha) - $\log(\sigma)$  
correlation found independently for LIRGs and ULIRGs is remarkable. Because the
average infrared luminosity ($\log(L_{ir}/L_\odot)$) of the ULIRG
sample is 12.24, this extends the previous result found for class 1
and class 2 LIRGs, where a similar linear relation exists for the
entire LIRG luminosity  range, independent of the luminosity beam
selected. Although the number of ULIRGs in the MAC06 sample is 
small, the combined results suggest a common
  relation between the excitation and dynamical properties of the 
ionized gas in interacting and merging (U)LIRGs over the entire
infrared luminosity range of $\log(L_{ir}/L_\odot) = 11.1- 12.3$
covered by these 
two samples. This relation is best traced by the direct
proportionality (slope of $\sim$1.0) between $\log$(\oha) and
$\log(\sigma)$  (see Tables \ref{pearsonvimos} and
\ref{pearsonvimos_integral} for the specific slopes).    
A larger sample of about 20 ULIRGs with available optical
integral field spectroscopy  
is currently studied (Garc\'{\i}a-Mar\'{\i}n et al. in preparation) to
further investigate the reality of the relation for both LIRGs
  and ULIRGs and its physical interpretation.

   \begin{table*}
\centering
      \caption[]{Linear Pearson correlation coefficients $r$ and
      linear fits for all C1 and C2 LIRGs and ULIRGs
      \label{pearsonvimos_integral}} 
              \begin{tabular}{ccccccccccccc}
            \hline
            \noalign{\smallskip}
Group & 
\multicolumn{2}{c}{[N\,\textsc{ii}]$\lambda$6584/H$\alpha$} &
\multicolumn{2}{c}{[S\,\textsc{ii}]$\lambda$6717,6731/H$\alpha$} &
\multicolumn{2}{c}{[O\,\textsc{i}]$\lambda$6300/H$\alpha$}\\
& r & A+Bx &  r & A+Bx & r & A+Bx \\
\hline
ULIRGs MAC06                     & 0.61 & -1.99 + 0.82 x & 0.51 & -1.36 + 0.50 x & 0.71 & -3.15 + 1.11 x\\
LIRGs C1+C2                      & 0.38 & -0.96 + 0.35 x & 0.41 & -1.09 + 0.36 x & 0.45 & -2.74 + 0.79 x\\
            \hline
         \end{tabular}
  \end{table*}

\section{Conclusions}

The two-dimensional ionization structure of the extended (few to several kpc) 
ionized gas in a representative sample of 32 low-$z$ LIRGs
(i.e. $\log(L_{ir}/L_\odot)=11.00-12.00$ luminosity range) was  
investigated with the VIMOS integral field spectrograph. The sample covers 
isolated galaxies, as well as interacting galaxies and systems in 
an advanced stage of the merger. This paper investigates the nature
and origin of the main ionization mechanisms operating in the 
extra-nuclear regions of these systems based on several thousands of
independent measurements of the emission line ratios (up to twenty
24\,000 for \nha) and velocity dispersions. 
 
The present study is part of a larger project devoted to the study of
the two-dimensional structure for the stars and ionized gas as well as
  its kinematics and ionization conditions in
representative samples of LIRGs and ULIRGs using optical IFS. 
The main results of this study can be summarized as follows:

1. The distribution of the \nha\  line ratio does not show any
   significant variations with the interaction class, with most regions
   presenting a ratio typical of \ion{H}{ii} regions. The \sha\ and \oha\
   line ratios do however show a change in their distribution
   with an extension 
   towards higher excitation (i.e. LINER-like excitation) for galaxies
   classified as interacting pairs and advanced mergers. This change
   is more pronounced for the \oha\ ratio.

2. There is an anti-correlation between the ionization
  degree and the \ha\ surface brightness independently of the interaction
  type and similar to what occurs in our Galaxy or in the so-called DIG
  in other spiral galaxies. Most of the observed line ratios are
  similar to those found in \ion{H}{ii} regions in our Galaxy, but
  there is a relatively large percentage of line ratios similar to
  those for DIGs.

3. The nature of the ionization sources was investigated comparing the
   measured \sha\ vs. \nha{} and \oha{} vs. \nha{} line
   ratios with the predictions of ionization due to stars
   (\ion{H}{ii} regions), TML, shocks (DIGs) and power-law
   (AGN) spectra. Turbulent Mixing Layers do not seem to play a major role in the ionization of the
     extra-nuclear regions. Line ratios in LIRGs classified as 
   isolated can mostly be explained as caused by ionization due to
   young stars. On the other hand, the
   ionization in a large fraction of the regions in systems with
   some degree  of interaction cannot be due to
   stars but is better explained by high
   velocity shocks. This is particularly   
   evident when using the best shock tracer, i.e. the \oha\ vs. \nha\
   diagram. Independently of the ionization mechanisms, only models
   with metallicity between solar and twice solar are able to explain
   the observed line ratios.
   
4. Local velocity dispersions  increase with the interaction degree, with medians of 37, 46, and 51~km~$^{-1}$ for class 0, 1, and 2 respectively, and are higher than those typically found in normal spirals \citep[$\sim20-30$~km~s$^{-1}$,][]{epi10}. This indicates that the ionized ISM in LIRGs is dynamically hotter than the quiescent ISM of normal galaxies due to strong shocks produced by tidal forces and stellar winds associated to the nuclear starburst.
     
5. There is a positive relation between the degree of
   excitation (as traced by 
   the emission line ratios) and the velocity dispersion of the
   ionized gas in LIRGs classified as interacting systems, 
   and mergers, while this relation is not observed in
   isolated systems. This relation is better seen when using
   the \oha\ and \sha\ line ratios, and supports the scenario where
   the relevance of shocks as ionizing sources in the
     extranuclear extended regions of LIRGs increases 
   when there is some degree of interaction.  

6. The relation between the degree of excitation and the velocity
   dispersion of the ionized gas in interacting and merging LIRGs does
   not clearly improve with the infrared luminosity (i.e. star formation rate)
   of the systems. Thus the interaction process itself rather than 
   superwinds caused by the star formation seems to be the main origin
   of the shocks in the extended extra-nuclear regions,
   assuming that the release of energy into the ISM is
     proportional to the SFR.  
   This result is still compatible with stellar
   superwinds in the internal regions of these systems, and/or along
   certain preferential directions associated with AGN-related outflows.

7. A comparison between the sub-sample of interacting/merging  
   LIRGs and a small sample of ULIRGs suggests the
   existence of a common positive $\log$(\oha) -
   $\log(\sigma)$ relation.   
   If confirmed, these results will provide further evidence for the
   tidal origin of shocks in these galaxies over the entire LIRG and
   ULIRG luminosity range. A study with a larger sample of ULIRGs is
   under way to confirm the result. 

\begin{acknowledgements}

We thank J. Alfonso-Garz\'on for her help in the
   initial stages of this project.
We also thank the anonymous referee for his/her careful and detailed
review of the manuscript that helped to greatly improve this paper.
Based on observations collected at the 
   European Organisation for Astronomical Research in the Southern
   Hemisphere, Chile (ESO Programs 076.B-0479(A), 078.B-0072(A) and
   081.B-0108(A)).
AM-I is grateful for the hospitality of the Instituto de Estructura de
la Materia where part of this work was performed. 
This paper uses the plotting package \texttt{jmaplot}, developed by Jes\'us
Ma\'{\i}z-Apell\'aniz,
\texttt{http://dae45.iaa.csic.es:8080/$\sim$jmaiz/software}. This 
research made use of the NASA/IPAC Extragalactic 
Database (NED), which is operated by the Jet Propulsion Laboratory, California
Institute of Technology, under contract with the National Aeronautics and Space
Administration.

This work has been supported by the Spanish Ministry for Education and
Science under grants PNE2005-01480 and ESP2007-65475-C02-01. 
AM-I is supported by the Spanish Ministry of 
Science and Innovation (MICINN) under the program "Specialization in
International Organisms", ref. ES2006-0003. 
MG-M is supported by the German federal department for education and
research (BMBF) under the project numbers: 50OS0502 \& 50OS0801.

\end{acknowledgements}


\bibliography{mybib}{}

\begin{thebibliography}{85}
\expandafter\ifx\csname natexlab\endcsname\relax\def\natexlab#1{#1}\fi

\bibitem[{{Allen} {et~al.}(2008){Allen}, {Groves}, {Dopita}, {Sutherland}, \&
  {Kewley}}]{all08}
{Allen}, M.~G., {Groves}, B.~A., {Dopita}, M.~A., {Sutherland}, R.~S., \&
  {Kewley}, L.~J. 2008, \apjs, 178, 20

\bibitem[{{Alonso-Herrero} {et~al.}(2009){Alonso-Herrero},
  {Garc{\'{\i}}a-Mar{\'{\i}}n}, {Monreal-Ibero}, {Colina}, {Arribas},
  {Alfonso-Garz{\'o}n}, \& {Labiano}}]{alo09}
{Alonso-Herrero}, A., {Garc{\'{\i}}a-Mar{\'{\i}}n}, M., {Monreal-Ibero}, A.,
  {et~al.} 2009, \aap, 506, 1541

\bibitem[{{Alonso-Herrero} {et~al.}(2010){Alonso-Herrero},
  {Garc{\'{\i}}a-Mar{\'{\i}}n}, {Rodriguez-Zaurin}, {Monreal-Ibero}, {Colina},
  \& {Arribas}}]{alo10}
{Alonso-Herrero}, A., {Garc{\'{\i}}a-Mar{\'{\i}}n}, M., {Rodriguez-Zaurin}, J.,
  {et~al.} 2010, summitted

\bibitem[{{Alonso-Herrero} {et~al.}(2006){Alonso-Herrero}, {Rieke}, {Rieke},
  {Colina}, {P{\'e}rez-Gonz{\'a}lez}, \& {Ryder}}]{alo06}
{Alonso-Herrero}, A., {Rieke}, G.~H., {Rieke}, M.~J., {et~al.} 2006, \apj, 650,
  835

\bibitem[{{Alonso-Herrero} {et~al.}(2002){Alonso-Herrero}, {Rieke}, {Rieke}, \&
  {Scoville}}]{alo02}
{Alonso-Herrero}, A., {Rieke}, G.~H., {Rieke}, M.~J., \& {Scoville}, N.~Z.
  2002, \aj, 124, 166

\bibitem[{{Armus} {et~al.}(1989){Armus}, {Heckman}, \& {Miley}}]{arm89}
{Armus}, L., {Heckman}, T.~M., \& {Miley}, G.~K. 1989, \apj, 347, 727

\bibitem[{{Arribas} {et~al.}(2004){Arribas}, {Bushouse}, {Lucas}, {Colina}, \&
  {Borne}}]{arr04}
{Arribas}, S., {Bushouse}, H., {Lucas}, R.~A., {Colina}, L., \& {Borne}, K.~D.
  2004, \aj, 127, 2522

\bibitem[{{Arribas} {et~al.}(1998){Arribas}, {Carter}, {Cavaller}, {del Burgo},
  {Edwards}, {Fuentes}, {Garcia}, {Herreros}, {Jones}, {Mediavilla}, {Pi},
  {Pollacco}, {Rasilla}, {Rees}, \& {Sosa}}]{arr98}
{Arribas}, S., {Carter}, D., {Cavaller}, L., {et~al.} 1998, in Society of
  Photo-Optical Instrumentation Engineers (SPIE) Conference Series, Vol. 3355,
  Society of Photo-Optical Instrumentation Engineers (SPIE) Conference Series,
  ed. S.~{D'Odorico}, 821--827

\bibitem[{{Arribas} {et~al.}(2001){Arribas}, {Colina}, \& {Clements}}]{arr01}
{Arribas}, S., {Colina}, L., \& {Clements}, D. 2001, \apj, 560, 160

\bibitem[{{Arribas} {et~al.}(2008){Arribas}, {Colina}, {Monreal-Ibero},
  {Alfonso}, {Garc{\'{\i}}a-Mar{\'{\i}}n}, \& {Alonso-Herrero}}]{arr08}
{Arribas}, S., {Colina}, L., {Monreal-Ibero}, A., {et~al.} 2008, \aap, 479, 687

\bibitem[{{Asplund} {et~al.}(2004){Asplund}, {Grevesse}, {Sauval}, {Allende
  Prieto}, \& {Kiselman}}]{asp04}
{Asplund}, M., {Grevesse}, N., {Sauval}, A.~J., {Allende Prieto}, C., \&
  {Kiselman}, D. 2004, \aap, 417, 751

\bibitem[{{Baldwin} {et~al.}(1981){Baldwin}, {Phillips}, \&
  {Terlevich}}]{bal81}
{Baldwin}, J.~A., {Phillips}, M.~M., \& {Terlevich}, R. 1981, \pasp, 93, 5

\bibitem[{{Bedregal} {et~al.}(2009){Bedregal}, {Colina}, {Alonso-Herrero}, \&
  {Arribas}}]{bed09}
{Bedregal}, A.~G., {Colina}, L., {Alonso-Herrero}, A., \& {Arribas}, S. 2009,
  \apj, 698, 1852

\bibitem[{{Bingham} {et~al.}(1994){Bingham}, {Gellatly}, {Jenkins}, \&
  {Worswick}}]{bin94}
{Bingham}, R.~G., {Gellatly}, D.~W., {Jenkins}, C.~R., \& {Worswick}, S.~P.
  1994, in Society of Photo-Optical Instrumentation Engineers (SPIE) Conference
  Series, Vol. 2198, Society of Photo-Optical Instrumentation Engineers (SPIE)
  Conference Series, ed. D.~L. {Crawford} \& E.~R. {Craine}, 56--64

\bibitem[{{Blanc} {et~al.}(2009){Blanc}, {Heiderman}, {Gebhardt}, {Evans}, \&
  {Adams}}]{bla09}
{Blanc}, G.~A., {Heiderman}, A., {Gebhardt}, K., {Evans}, N.~J., \& {Adams}, J.
  2009, \apj, 704, 842

\bibitem[{{Borne} {et~al.}(2000){Borne}, {Bushouse}, {Lucas}, \&
  {Colina}}]{bor00}
{Borne}, K.~D., {Bushouse}, H., {Lucas}, R.~A., \& {Colina}, L. 2000, \apjl,
  529, L77

\bibitem[{{Bushouse} {et~al.}(2002){Bushouse}, {Borne}, {Colina}, {Lucas},
  {Rowan-Robinson}, {Baker}, {Clements}, {Lawrence}, \& {Oliver}}]{bus02}
{Bushouse}, H.~A., {Borne}, K.~D., {Colina}, L., {et~al.} 2002, \apjs, 138, 1

\bibitem[{{Clements} {et~al.}(1996){Clements}, {Sutherland}, {McMahon}, \&
  {Saunders}}]{cle96}
{Clements}, D.~L., {Sutherland}, W.~J., {McMahon}, R.~G., \& {Saunders}, W.
  1996, \mnras, 279, 477

\bibitem[{{Colina} {et~al.}(2005){Colina}, {Arribas}, \&
  {Monreal-Ibero}}]{col05}
{Colina}, L., {Arribas}, S., \& {Monreal-Ibero}, A. 2005, \apj, 621, 725

\bibitem[{{Colina} {et~al.}(2003){Colina}, {Gonz\'alez Delgado}, {Mas-Hesse},
  {Leitherer}, \& {Jim{\'e}nez Bail{\'o}n}}]{col03}
{Colina}, L., {Gonz\'alez Delgado}, R., {Mas-Hesse}, J.~M., {Leitherer}, C., \&
  {Jim{\'e}nez Bail{\'o}n}, E. 2003, \apj, 582, 1269

\bibitem[{{Colina} {et~al.}(1991){Colina}, {Lipari}, \& {Macchetto}}]{col91}
{Colina}, L., {Lipari}, S., \& {Macchetto}, F. 1991, \apj, 379, 113

\bibitem[{{Collins} \& {Rand}(2001)}]{col01d}
{Collins}, J.~A. \& {Rand}, R.~J. 2001, \apj, 551, 57

\bibitem[{{Corbett} {et~al.}(2003){Corbett}, {Kewley}, {Appleton},
  {Charmandaris}, {Dopita}, {Heisler}, {Norris}, {Zezas}, \& {Marston}}]{cor03}
{Corbett}, E.~A., {Kewley}, L., {Appleton}, P.~N., {et~al.} 2003, \apj, 583,
  670

\bibitem[{{Cui} {et~al.}(2001){Cui}, {Xia}, {Deng}, {Mao}, \& {Zou}}]{cui01}
{Cui}, J., {Xia}, X.-Y., {Deng}, Z.-G., {Mao}, S., \& {Zou}, Z.-L. 2001, \aj,
  122, 63

\bibitem[{{Dopita} {et~al.}(2006){Dopita}, {Fischera}, {Sutherland}, {Kewley},
  {Leitherer}, {Tuffs}, {Popescu}, {van Breugel}, \& {Groves}}]{dop06}
{Dopita}, M.~A., {Fischera}, J., {Sutherland}, R.~S., {et~al.} 2006, \apjs,
  167, 177

\bibitem[{{Dopita} \& {Sutherland}(1995)}]{dop95}
{Dopita}, M.~A. \& {Sutherland}, R.~S. 1995, \apj, 455, 468

\bibitem[{{Eisenhauer} {et~al.}(2003){Eisenhauer}, {Abuter}, {Bickert},
  {Biancat-Marchet}, {Bonnet}, {Brynnel}, {Conzelmann}, {Delabre}, {Donaldson},
  {Farinato}, {Fedrigo}, {Genzel}, {Hubin}, {Iserlohe}, {Kasper},
  {Kissler-Patig}, {Monnet}, {Roehrle}, {Schreiber}, {Stroebele}, {Tecza},
  {Thatte}, \& {Weisz}}]{eis03}
{Eisenhauer}, F., {Abuter}, R., {Bickert}, K., {et~al.} 2003, in Society of
  Photo-Optical Instrumentation Engineers (SPIE) Conference Series, Vol. 4841,
  Society of Photo-Optical Instrumentation Engineers (SPIE) Conference Series,
  ed. M.~{Iye} \& A.~F.~M. {Moorwood}, 1548--1561

\bibitem[{{Elbaz} {et~al.}(2002){Elbaz}, {Cesarsky}, {Chanial}, {Aussel},
  {Franceschini}, {Fadda}, \& {Chary}}]{elb02}
{Elbaz}, D., {Cesarsky}, C.~J., {Chanial}, P., {et~al.} 2002, \aap, 384, 848

\bibitem[{{Epinat} {et~al.}(2010){Epinat}, {Amram}, {Balkowski}, \&
  {Marcelin}}]{epi10}
{Epinat}, B., {Amram}, P., {Balkowski}, C., \& {Marcelin}, M. 2010, \mnras,
  401, 2113

\bibitem[{{Evans} {et~al.}(2002){Evans}, {Mazzarella}, {Surace}, \&
  {Sanders}}]{eva02}
{Evans}, A.~S., {Mazzarella}, J.~M., {Surace}, J.~A., \& {Sanders}, D.~B. 2002,
  \apj, 580, 749

\bibitem[{{Farrah} {et~al.}(2007){Farrah}, {Bernard-Salas}, {Spoon}, {Soifer},
  {Armus}, {Brandl}, {Charmandaris}, {Desai}, {Higdon}, {Devost}, \&
  {Houck}}]{far07}
{Farrah}, D., {Bernard-Salas}, J., {Spoon}, H.~W.~W., {et~al.} 2007, \apj, 667,
  149

\bibitem[{{Garc{\'{\i}}a-Mar{\'{\i}}n}
  {et~al.}(2006){Garc{\'{\i}}a-Mar{\'{\i}}n}, {Colina}, {Arribas},
  {Alonso-Herrero}, \& {Mediavilla}}]{gar06}
{Garc{\'{\i}}a-Mar{\'{\i}}n}, M., {Colina}, L., {Arribas}, S.,
  {Alonso-Herrero}, A., \& {Mediavilla}, E. 2006, \apj, 650, 850

\bibitem[{{Garc{\'{\i}}a-Mar{\'{\i}}n}
  {et~al.}(2009){Garc{\'{\i}}a-Mar{\'{\i}}n}, {Colina}, {Arribas}, \&
  {Monreal-Ibero}}]{gar09}
{Garc{\'{\i}}a-Mar{\'{\i}}n}, M., {Colina}, L., {Arribas}, S., \&
  {Monreal-Ibero}, A. 2009, \aap, 505, 1319

\bibitem[{{Genzel} {et~al.}(1998){Genzel}, {Lutz}, {Sturm}, {Egami}, {Kunze},
  {Moorwood}, {Rigopoulou}, {Spoon}, {Sternberg}, {Tacconi-Garman}, {Tacconi},
  \& {Thatte}}]{gen98}
{Genzel}, R., {Lutz}, D., {Sturm}, E., {et~al.} 1998, \apj, 498, 579

\bibitem[{{Groves} {et~al.}(2004){Groves}, {Dopita}, \& {Sutherland}}]{gro04}
{Groves}, B.~A., {Dopita}, M.~A., \& {Sutherland}, R.~S. 2004, \apjs, 153, 75

\bibitem[{{Heckman} {et~al.}(1990){Heckman}, {Armus}, \& {Miley}}]{hec90}
{Heckman}, T.~M., {Armus}, L., \& {Miley}, G.~K. 1990, \apjs, 74, 833

\bibitem[{{Heckman} {et~al.}(2000){Heckman}, {Lehnert}, {Strickland}, \&
  {Armus}}]{hec00}
{Heckman}, T.~M., {Lehnert}, M.~D., {Strickland}, D.~K., \& {Armus}, L. 2000,
  \apjs, 129, 493

\bibitem[{{Hinz} \& {Rieke}(2006)}]{hin06}
{Hinz}, J.~L. \& {Rieke}, G.~H. 2006, \apj, 646, 872

\bibitem[{{Kauffmann} {et~al.}(2003){Kauffmann}, {Heckman}, {Tremonti},
  {Brinchmann}, {Charlot}, {White}, {Ridgway}, {Brinkmann}, {Fukugita}, {Hall},
  {Ivezi{\'c}}, {Richards}, \& {Schneider}}]{kau03}
{Kauffmann}, G., {Heckman}, T.~M., {Tremonti}, C., {et~al.} 2003, \mnras, 346,
  1055

\bibitem[{{Kennicutt}(1984)}]{ken84}
{Kennicutt}, Jr., R.~C. 1984, \apj, 287, 116

\bibitem[{{Kennicutt}(1998)}]{ken98}
{Kennicutt}, Jr., R.~C. 1998, \araa, 36, 189

\bibitem[{{Kewley} {et~al.}(2001{\natexlab{a}}){Kewley}, {Dopita},
  {Sutherland}, {Heisler}, \& {Trevena}}]{kew01a}
{Kewley}, L.~J., {Dopita}, M.~A., {Sutherland}, R.~S., {Heisler}, C.~A., \&
  {Trevena}, J. 2001{\natexlab{a}}, \apj, 556, 121

\bibitem[{{Kewley} {et~al.}(2006){Kewley}, {Groves}, {Kauffmann}, \&
  {Heckman}}]{kew06}
{Kewley}, L.~J., {Groves}, B., {Kauffmann}, G., \& {Heckman}, T. 2006, \mnras,
  372, 961

\bibitem[{{Kewley} {et~al.}(2001{\natexlab{b}}){Kewley}, {Heisler}, {Dopita},
  \& {Lumsden}}]{kew01b}
{Kewley}, L.~J., {Heisler}, C.~A., {Dopita}, M.~A., \& {Lumsden}, S.
  2001{\natexlab{b}}, \apjs, 132, 37

\bibitem[{{Kim} {et~al.}(1995){Kim}, {Sanders}, {Veilleux}, {Mazzarella}, \&
  {Soifer}}]{kim95}
{Kim}, D.-C., {Sanders}, D.~B., {Veilleux}, S., {Mazzarella}, J.~M., \&
  {Soifer}, B.~T. 1995, \apjs, 98, 129

\bibitem[{{Le Floc'h} {et~al.}(2005){Le Floc'h}, {Papovich}, {Dole}, {Bell},
  {Lagache}, {Rieke}, {Egami}, {P{\'e}rez-Gonz{\'a}lez}, {Alonso-Herrero},
  {Rieke}, {Blaylock}, {Engelbracht}, {Gordon}, {Hines}, {Misselt}, {Morrison},
  \& {Mould}}]{lef05}
{Le Floc'h}, E., {Papovich}, C., {Dole}, H., {et~al.} 2005, \apj, 632, 169

\bibitem[{{LeF\`evre} {et~al.}(2003){LeF\`evre}, {Saisse}, {Mancini},
  {Brau-Nogue}, {Caputi}, {Castinel}, {D'Odorico}, {Garilli}, {Kissler-Patig},
  {Lucuix}, {Mancini}, {Pauget}, {Sciarretta}, {Scodeggio}, {Tresse}, \&
  {Vettolani}}]{lef03}
{LeF\`evre}, O., {Saisse}, M., {Mancini}, D., {et~al.} 2003, in Society of
  Photo-Optical Instrumentation Engineers (SPIE) Conference Series, Vol. 4841,
  Society of Photo-Optical Instrumentation Engineers (SPIE) Conference Series,
  ed. M.~{Iye} \& A.~F.~M. {Moorwood}, 1670--1681

\bibitem[{{Lehnert} \& {Heckman}(1996)}]{leh96}
{Lehnert}, M.~D. \& {Heckman}, T.~M. 1996, \apj, 462, 651

\bibitem[{{Leitherer} {et~al.}(1999){Leitherer}, {Schaerer}, {Goldader},
  {Delgado}, {Robert}, {Kune}, {de Mello}, {Devost}, \& {Heckman}}]{lei99}
{Leitherer}, C., {Schaerer}, D., {Goldader}, J.~D., {et~al.} 1999, \apjs, 123,
  3

\bibitem[{{Lonsdale} {et~al.}(2006){Lonsdale}, {Farrah}, \& {Smith}}]{lon06}
{Lonsdale}, C.~J., {Farrah}, D., \& {Smith}, H.~E. 2006, {Ultraluminous
  Infrared Galaxies}, ed. J.~W. {Mason} (Springer Verlag), 285--+

\bibitem[{{Madsen} {et~al.}(2006){Madsen}, {Reynolds}, \& {Haffner}}]{mad06}
{Madsen}, G.~J., {Reynolds}, R.~J., \& {Haffner}, L.~M. 2006, \apj, 652, 401

\bibitem[{{Martin}(2005)}]{mar05}
{Martin}, C.~L. 2005, \apj, 621, 227

\bibitem[{{Martin}(2006)}]{mar06}
{Martin}, C.~L. 2006, \apj, 647, 222

\bibitem[{{Mathis}(2000)}]{mat00}
{Mathis}, J.~S. 2000, \apj, 544, 347

\bibitem[{{McDowell} {et~al.}(2003){McDowell}, {Clements}, {Lamb}, {Shaked},
  {Hearn}, {Colina}, {Mundell}, {Borne}, {Baker}, \& {Arribas}}]{mcd03}
{McDowell}, J.~C., {Clements}, D.~L., {Lamb}, S.~A., {et~al.} 2003, \apj, 591,
  154

\bibitem[{{Mihos} \& {Hernquist}(1996)}]{mih96}
{Mihos}, J.~C. \& {Hernquist}, L. 1996, \apj, 464, 641

\bibitem[{{Miller} \& {Veilleux}(2003)}]{mil03}
{Miller}, S.~T. \& {Veilleux}, S. 2003, \apj, 592, 79

\bibitem[{{Monreal-Ibero} {et~al.}(2006){Monreal-Ibero}, {Arribas}, \&
  {Colina}}]{mon06}
{Monreal-Ibero}, A., {Arribas}, S., \& {Colina}, L. 2006, \apj, 637, 138

\bibitem[{{Moshir} \& {et al.}(1990)}]{mos90}
{Moshir}, M. \& {et al.} 1990, in IRAS Faint Source Catalogue, version 2.0
  (1990), 0--+

\bibitem[{{Naab} {et~al.}(2006){Naab}, {Jesseit}, \& {Burkert}}]{naa06}
{Naab}, T., {Jesseit}, R., \& {Burkert}, A. 2006, \mnras, 372, 839

\bibitem[{{Nardini} {et~al.}(2008){Nardini}, {Risaliti}, {Salvati}, {Sani},
  {Imanishi}, {Marconi}, \& {Maiolino}}]{nar08}
{Nardini}, E., {Risaliti}, G., {Salvati}, M., {et~al.} 2008, \mnras, 385, L130

\bibitem[{{P{\'e}rez-Gonz{\'a}lez} {et~al.}(2005){P{\'e}rez-Gonz{\'a}lez},
  {Rieke}, {Egami}, {Alonso-Herrero}, {Dole}, {Papovich}, {Blaylock}, {Jones},
  {Rieke}, {Rigby}, {Barmby}, {Fazio}, {Huang}, \& {Martin}}]{per05}
{P{\'e}rez-Gonz{\'a}lez}, P.~G., {Rieke}, G.~H., {Egami}, E., {et~al.} 2005,
  \apj, 630, 82

\bibitem[{{Reynolds} {et~al.}(1999){Reynolds}, {Haffner}, \& {Tufte}}]{rey99}
{Reynolds}, R.~J., {Haffner}, L.~M., \& {Tufte}, S.~L. 1999, \apjl, 525, L21

\bibitem[{{Risaliti} {et~al.}(2006){Risaliti}, {Maiolino}, {Marconi}, {Sani},
  {Berta}, {Braito}, {Ceca}, {Franceschini}, \& {Salvati}}]{ris06}
{Risaliti}, G., {Maiolino}, R., {Marconi}, A., {et~al.} 2006, \mnras, 365, 303

\bibitem[{{Rodr{\'{\i}}guez-Zaur{\'{\i}}n}
  {et~al.}(2010){Rodr{\'{\i}}guez-Zaur{\'{\i}}n}, {Arribas}, {Monreal-Ibero},
  {Colina}, {Alonso-Herrero}, \& {Alfonso-Garz\'on}}]{rod10}
{Rodr{\'{\i}}guez-Zaur{\'{\i}}n}, J., {Arribas}, S., {Monreal-Ibero}, A.,
  {et~al.} 2010, in prep.

\bibitem[{{Roth} {et~al.}(2005){Roth}, {Kelz}, {Fechner}, {Hahn}, {Bauer},
  {Becker}, {B{\"o}hm}, {Christensen}, {Dionies}, {Paschke}, {Popow}, {Wolter},
  {Schmoll}, {Laux}, \& {Altmann}}]{rot05}
{Roth}, M.~M., {Kelz}, A., {Fechner}, T., {et~al.} 2005, \pasp, 117, 620

\bibitem[{{Rupke} {et~al.}(2002){Rupke}, {Veilleux}, \& {Sanders}}]{rup02}
{Rupke}, D.~S., {Veilleux}, S., \& {Sanders}, D.~B. 2002, \apj, 570, 588

\bibitem[{{Rupke} {et~al.}(2005{\natexlab{a}}){Rupke}, {Veilleux}, \&
  {Sanders}}]{rup05b}
{Rupke}, D.~S., {Veilleux}, S., \& {Sanders}, D.~B. 2005{\natexlab{a}}, \apjs,
  160, 87

\bibitem[{{Rupke} {et~al.}(2005{\natexlab{b}}){Rupke}, {Veilleux}, \&
  {Sanders}}]{rup05a}
{Rupke}, D.~S., {Veilleux}, S., \& {Sanders}, D.~B. 2005{\natexlab{b}}, \apjs,
  160, 115

\bibitem[{{Rupke} {et~al.}(2008){Rupke}, {Veilleux}, \& {Baker}}]{rup08}
{Rupke}, D.~S.~N., {Veilleux}, S., \& {Baker}, A.~J. 2008, \apj, 674, 172

\bibitem[{{Sanders} \& {Ishida}(2004)}]{san04}
{Sanders}, D. \& {Ishida}, C. 2004, in Astronomical Society of the Pacific
  Conference Series, Vol. 320, The Neutral ISM in Starburst Galaxies, ed.
  S.~{Aalto}, S.~{Huttemeister}, \& A.~{Pedlar}, 230--+

\bibitem[{{Sanders} {et~al.}(2003){Sanders}, {Mazzarella}, {Kim}, {Surace}, \&
  {Soifer}}]{san03}
{Sanders}, D.~B., {Mazzarella}, J.~M., {Kim}, D.-C., {Surace}, J.~A., \&
  {Soifer}, B.~T. 2003, \aj, 126, 1607

\bibitem[{{Sanders} \& {Mirabel}(1996)}]{san96}
{Sanders}, D.~B. \& {Mirabel}, I.~F. 1996, \araa, 34, 749

\bibitem[{{Scoville} {et~al.}(2000){Scoville}, {Evans}, {Thompson}, {Rieke},
  {Hines}, {Low}, {Dinshaw}, {Surace}, \& {Armus}}]{sco00}
{Scoville}, N.~Z., {Evans}, A.~S., {Thompson}, R., {et~al.} 2000, \aj, 119, 991

\bibitem[{{Slavin} {et~al.}(1993){Slavin}, {Shull}, \& {Begelman}}]{sla93}
{Slavin}, J.~D., {Shull}, J.~M., \& {Begelman}, M.~C. 1993, \apj, 407, 83

\bibitem[{{Stasi{\'n}ska} {et~al.}(2006){Stasi{\'n}ska}, {Cid Fernandes},
  {Mateus}, {Sodr{\'e}}, \& {Asari}}]{sta06}
{Stasi{\'n}ska}, G., {Cid Fernandes}, R., {Mateus}, A., {Sodr{\'e}}, L., \&
  {Asari}, N.~V. 2006, \mnras, 371, 972

\bibitem[{{Tremonti} {et~al.}(2004){Tremonti}, {Heckman}, {Kauffmann},
  {Brinchmann}, {Charlot}, {White}, {Seibert}, {Peng}, {Schlegel}, {Uomoto},
  {Fukugita}, \& {Brinkmann}}]{tre04}
{Tremonti}, C.~A., {Heckman}, T.~M., {Kauffmann}, G., {et~al.} 2004, \apj, 613,
  898

\bibitem[{{V{\"a}is{\"a}nen} {et~al.}(2008{\natexlab{a}}){V{\"a}is{\"a}nen},
  {Mattila}, {Kniazev}, {Adamo}, {Efstathiou}, {Farrah}, {Johansson},
  {{\"O}stlin}, {Buckley}, {Burgh}, {Crause}, {Hashimoto}, {Lira}, {Loaring},
  {Nordsieck}, {Romero-Colmenero}, {Ryder}, {Still}, \& {Zijlstra}}]{vai08b}
{V{\"a}is{\"a}nen}, P., {Mattila}, S., {Kniazev}, A., {et~al.}
  2008{\natexlab{a}}, \mnras, 384, 886

\bibitem[{{V{\"a}is{\"a}nen} {et~al.}(2008{\natexlab{b}}){V{\"a}is{\"a}nen},
  {Ryder}, {Mattila}, \& {Kotilainen}}]{vai08a}
{V{\"a}is{\"a}nen}, P., {Ryder}, S., {Mattila}, S., \& {Kotilainen}, J.
  2008{\natexlab{b}}, \apjl, 689, L37

\bibitem[{{Veilleux} {et~al.}(1999){Veilleux}, {Kim}, \& {Sanders}}]{vei99}
{Veilleux}, S., {Kim}, D.-C., \& {Sanders}, D.~B. 1999, \apj, 522, 113

\bibitem[{{Veilleux} {et~al.}(2002){Veilleux}, {Kim}, \& {Sanders}}]{vei02}
{Veilleux}, S., {Kim}, D.-C., \& {Sanders}, D.~B. 2002, \apjs, 143, 315

\bibitem[{{Veilleux} {et~al.}(1995){Veilleux}, {Kim}, {Sanders}, {Mazzarella},
  \& {Soifer}}]{vei95}
{Veilleux}, S., {Kim}, D.-C., {Sanders}, D.~B., {Mazzarella}, J.~M., \&
  {Soifer}, B.~T. 1995, \apjs, 98, 171

\bibitem[{{Veilleux} \& {Osterbrock}(1987)}]{vei87}
{Veilleux}, S. \& {Osterbrock}, D.~E. 1987, \apjs, 63, 295

\bibitem[{{York} {et~al.}(2000){York}, {Adelman}, {Anderson}, {Anderson},
  {Annis}, {Bahcall}, {Bakken}, {Barkhouser}, {Bastian}, {Berman}, {Boroski},
  {Bracker}, {Briegel}, {Briggs}, {Brinkmann}, {Brunner}, {Burles}, {Carey},
  {Carr}, {Castander}, {Chen}, {Colestock}, {Connolly}, {Crocker}, {Csabai},
  {Czarapata}, {Davis}, {Doi}, {Dombeck}, {Eisenstein}, {Ellman}, {Elms},
  {Evans}, {Fan}, {Federwitz}, {Fiscelli}, {Friedman}, {Frieman}, {Fukugita},
  {Gillespie}, {Gunn}, {Gurbani}, {de Haas}, {Haldeman}, {Harris}, {Hayes},
  {Heckman}, {Hennessy}, {Hindsley}, {Holm}, {Holmgren}, {Huang}, {Hull},
  {Husby}, {Ichikawa}, {Ichikawa}, {Ivezi{\'c}}, {Kent}, {Kim}, {Kinney},
  {Klaene}, {Kleinman}, {Kleinman}, {Knapp}, {Korienek}, {Kron}, {Kunszt},
  {Lamb}, {Lee}, {Leger}, {Limmongkol}, {Lindenmeyer}, {Long}, {Loomis},
  {Loveday}, {Lucinio}, {Lupton}, {MacKinnon}, {Mannery}, {Mantsch}, {Margon},
  {McGehee}, {McKay}, {Meiksin}, {Merelli}, {Monet}, {Munn}, {Narayanan},
  {Nash}, {Neilsen}, {Neswold}, {Newberg}, {Nichol}, {Nicinski}, {Nonino},
  {Okada}, {Okamura}, {Ostriker}, {Owen}, {Pauls}, {Peoples}, {Peterson},
  {Petravick}, {Pier}, {Pope}, {Pordes}, {Prosapio}, {Rechenmacher}, {Quinn},
  {Richards}, {Richmond}, {Rivetta}, {Rockosi}, {Ruthmansdorfer}, {Sandford},
  {Schlegel}, {Schneider}, {Sekiguchi}, {Sergey}, {Shimasaku}, {Siegmund},
  {Smee}, {Smith}, {Snedden}, {Stone}, {Stoughton}, {Strauss}, {Stubbs},
  {SubbaRao}, {Szalay}, {Szapudi}, {Szokoly}, {Thakar}, {Tremonti}, {Tucker},
  {Uomoto}, {Vanden Berk}, {Vogeley}, {Waddell}, {Wang}, {Watanabe},
  {Weinberg}, {Yanny}, \& {Yasuda}}]{yor00}
{York}, D.~G., {Adelman}, J., {Anderson}, Jr., J.~E., {et~al.} 2000, \aj, 120,
  1579

\bibitem[{{Yuan} {et~al.}(2010){Yuan}, {Kewley}, \& {Sanders}}]{yua10}
{Yuan}, T.-T., {Kewley}, L.~J., \& {Sanders}, D.~B. 2010, \mnras, 709, 884

\end{thebibliography}
\bibliographystyle{./aa}




\onlfig{14}{
\begin{figure*}[!ht]
   \centering
\textbf{FULL VERSION at http://www.damir.iem.csic.es/extragalactic/publications/publications.html}\\
   \caption[ ]{Panel showing the images utilized for the morphological
     classification. The left column contains details about the size and
     scale for the displayed images, as well as relevant morphological
     features and final classification. The central and right columns
     display the \emph{Digital Sky Survey} and HST images,
     respectively. Orientation is north up, east to the left.
   \label{panel}}
\end{figure*}
}

\onlfig{14}{
\begin{figure*}[!ht]
   \centering
\textbf{FULL VERSION at http://www.damir.iem.csic.es/extragalactic/publications/publications.html}\\
\textbf{Fig. \ref{panel}.} continued.
\end{figure*}
}

\onlfig{14}{
\begin{figure*}[!ht]
   \centering
\textbf{FULL VERSION at http://www.damir.iem.csic.es/extragalactic/publications/publications.html}\\
\textbf{Fig. \ref{panel}.} continued.
\end{figure*}
}

\onlfig{14}{
\begin{figure*}[!ht]
   \centering
\textbf{FULL VERSION at http://www.damir.iem.csic.es/extragalactic/publications/publications.html}\\
\textbf{Fig. \ref{panel}.} continued.
\end{figure*}
}

\onlfig{14}{
\begin{figure*}[!ht]
   \centering
\textbf{FULL VERSION at http://www.damir.iem.csic.es/extragalactic/publications/publications.html}\\
\textbf{Fig. \ref{panel}.} continued.
\end{figure*}
}

\onlfig{14}{
\begin{figure*}[!ht]
   \centering
\textbf{FULL VERSION at http://www.damir.iem.csic.es/extragalactic/publications/publications.html}\\
\textbf{Fig. \ref{panel}.} continued.
\end{figure*}
}

\onlfig{14}{
\begin{figure*}[!ht]
   \centering
\textbf{FULL VERSION at http://www.damir.iem.csic.es/extragalactic/publications/publications.html}\\
\textbf{Fig. \ref{panel}.} continued.
\end{figure*}
}


\onlfig{15}{
\begin{figure*}[!ht]
   \centering
\textbf{FULL VERSION at http://www.damir.iem.csic.es/extragalactic/publications/publications.html}\\
   \caption[S\,\textsc{ii}$\lambda\lambda$6717,6731/H$\alpha$
   vs. N\,\textsc{ii}$\lambda$6584/H$\alpha$ for the individual systems]
   {[S\,\textsc{ii}]$\lambda\lambda$6717,6731/H$\alpha$
   vs. [N\,\textsc{ii}]$\lambda$6584/H$\alpha$ diagrams for the
   individual pointings. 
   \label{s2havsn2gha_indi}}
\end{figure*}
}

\onlfig{15}{
\begin{figure*}[!ht]
   \centering
\textbf{FULL VERSION at http://www.damir.iem.csic.es/extragalactic/publications/publications.html}\\
\textbf{Fig. \ref{s2havsn2gha_indi}.} continued.
\end{figure*}
}

\onlfig{16}{
\begin{figure*}[!ht]
 \centering
\textbf{FULL VERSION at http://www.damir.iem.csic.es/extragalactic/publications/publications.html}\\ \caption[O\,\textsc{i}$\lambda$6300/H$\alpha$
   vs. N\,\textsc{ii}$\lambda$6584/H$\alpha$ for the individual systems]
   {[O\,\textsc{i}]$\lambda$6300/H$\alpha$
   vs. [N\,\textsc{ii}]$\lambda$6584/H$\alpha$ diagrams for the
   individual pointings. 
   \label{o1havsn2gha_indi}}
\end{figure*}
}

\onlfig{16}{
\begin{figure*}[!ht]
 \centering
\textbf{FULL VERSION at http://www.damir.iem.csic.es/extragalactic/publications/publications.html}\\
\textbf{Fig. \ref{o1havsn2gha_indi}.} continued.
\end{figure*}
}


\onlfig{17}{
\begin{figure*}
\centering
\textbf{FULL VERSION at http://www.damir.iem.csic.es/extragalactic/publications/publications.html}\\
\caption{Line ratios vs. velocity dispersions relation for the
  individual pointings.}
\label{lirgs_indi}
\end{figure*}
}

\onlfig{17}{
\begin{figure*}[!ht]
   \centering
\textbf{FULL VERSION at http://www.damir.iem.csic.es/extragalactic/publications/publications.html}\\
\textbf{Fig. \ref{lirgs_indi}.} continued.
\end{figure*}
}

\onlfig{17}{
\begin{figure*}[!ht]
   \centering
\textbf{FULL VERSION at http://www.damir.iem.csic.es/extragalactic/publications/publications.html}\\
\textbf{Fig. \ref{lirgs_indi}.} continued.
\end{figure*}
}

\onlfig{17}{
\begin{figure*}[!ht]
   \centering
\textbf{FULL VERSION at http://www.damir.iem.csic.es/extragalactic/publications/publications.html}\\
\textbf{Fig. \ref{lirgs_indi}.} continued.
\end{figure*}
}

\onlfig{17}{
\begin{figure*}[!ht]
   \centering
\textbf{FULL VERSION at http://www.damir.iem.csic.es/extragalactic/publications/publications.html}\\
\textbf{Fig. \ref{lirgs_indi}.} continued.
\end{figure*}
}

\onlfig{17}{
\begin{figure*}[!ht]
   \centering
\textbf{FULL VERSION at http://www.damir.iem.csic.es/extragalactic/publications/publications.html}\\
\textbf{Fig. \ref{lirgs_indi}.} continued.
\end{figure*}
}

\onlfig{17}{
\begin{figure*}[!ht]
   \centering
\textbf{FULL VERSION at http://www.damir.iem.csic.es/extragalactic/publications/publications.html}\\
\textbf{Fig. \ref{lirgs_indi}.} continued.
\end{figure*}
}


\end{document}